\documentclass{jpp}
\usepackage{graphicx,color}
\usepackage[utf8]{inputenc}
\usepackage[T1]{fontenc}
\usepackage{amsmath}
\usepackage{lipsum}
\usepackage{upgreek}
\usepackage{textgreek}
\usepackage{gensymb}
\usepackage{hyperref}
\DeclareMathOperator\sinc{sinc}
\DeclareMathOperator\erf{erf}
\usepackage[dvipsnames]{xcolor}
\usepackage{bm}
\usepackage{gensymb}
\usepackage{mhchem}
\usepackage{textcase}

\shorttitle{Interaction of a Tightly Focused Single-Cycle Ultraintense Laser with a Solid}
\title{Gamma-Ray Flash in the Interaction of a Tightly Focused Single-Cycle Ultraintense Laser Pulse with a Solid Target}

\shortauthor{P. Hadjisolomou and others}
\author{P. Hadjisolomou\aff{1}
  \corresp{\email{Prokopis.Hadjisolomou@eli-beams.eu}},
  T. M. Jeong\aff{1},
  P. Valenta\aff{1,2},
  D. Kolenaty\aff{1},
  R. Versaci\aff{1},
  V. Olšovcová\aff{1},
  C. P. Ridgers\aff{3},
  \and S. V. Bulanov\aff{1,4}
}

\affiliation{
\aff{1} ELI Beamlines Centre, Institute of Physics, Czech Academy of Sciences, Za Radnicí 835, 25241 Dolní Břežany, Czech Republic
\aff{2} Faculty of Nuclear Sciences and Physical Engineering, Czech Technical University in Prague, Brehova 7, Prague 11519, Czech Republic
\aff{3} York Plasma Institute, Department of Physics, University of York, Heslington, York, North Yorkshire YO10 5DD, UK
\aff{4} National Institutes for Quantum and Radiological Science and Technology (QST), Kansai Photon Science Institute, 8-1-7 Umemidai, Kizugawa, Kyoto 619-0215, Japan
}

\begin{document}

\maketitle

\begin{abstract}
We employ the $\lambda^3$ regime where a near-single-cycle laser pulse is tightly focused, thus providing the highest possible intensity for the minimal energy at a certain laser power. The quantum electrodynamics processes in the course of the interaction of the ultraintense laser with a solid target are studied via three-dimensional particle-in-cell simulations, revealing the generation of copious \textgamma-photons and electron-positron pairs. The parametric study on the laser polarisation, target thickness and electron number density shows that the radially polarised laser provides the optimal regime for \textgamma-photon generation. By varying the laser power in the range of 1 to 300 petawatt we find the scaling of the laser to \textgamma-photon energy conversion efficiency. The laser-generated \textgamma-photon interaction with a high-Z target is further studied by using Monte Carlo simulations revealing further electron-positron pair generation and radioactive nuclides creation.
\end{abstract}


\section{Introduction} \label{Introduction}

\par The invention of the Chirped Pulse Amplification (CPA) technique \citep{1985_StricklandD} in mid-80's allowed the rapid growth of the laser power beyond the terawatt (TW) level. The petawatt (PW) threshold was exceeded at the end of 20th century\citep{1999_Perry}. Currently, the record power is for the ELI-NP $10 \kern0.2em \mathrm{PW}$ laser \citep{2020_TanakaKA}, with another $10 \kern0.2em \mathrm{PW}$ laser near completion in ELI-Beamlies. Current worldwide activities on PW laser systems and further envisions to attain ${>} \kern0.1em 100 \kern0.2em \mathrm{PW}$ lasers are summarized in \citep{2019_DansonC, 2021_LiZ}.  

\par Since the laser power increases by either increasing the energy or reducing the pulse duration, a single-cycle pulse was proposed \citep{2002_MourouG, 2006_BulanovSS, 2013_VoroninAA}. Post-compression of the CPA systems leads to near-single-cycle pulses by self-phase modulation in hollow-core fibres, although the energy is in the millijoule level \citep{2014_BohleF, 2020_OuilleM}. A second technique producing near-single-cycle pulses is the Optical Parametric CPA, by which a $4.5 \kern0.2em \mathrm{fs}$, $16 \kern0.2em \mathrm{TW}$ pulse is reported \citep{2017_Rivas}. Reducing the pulse duration is the primary goal of ELI-ALPS, where a $17 \kern0.2em \mathrm{fs}$, $2 \kern0.2em \mathrm{PW}$ laser is under development \citep{2019_Osvay}. Thus, at a given laser power, reduction of the pulse duration leads to a linear reduction of the energy, consequently the minimum laser energy for a single-cycle pulse.

\par However, it is most desired to reach the highest laser intensity rather than power. The quadratic dependency of the intensity on the inverse of the focal spot radius points on emphasizing for a reduced focal spot. More than two decades ago, a theoretical estimation of the minimum focal spot diameter \citep{1998_Sales} suggests a value of $4 \upi^{-2} \lambda$, where $\lambda$ is the laser wavelength. A vectorial diffraction approach was adopted \citep{1959_RichardsB, 2010_April} to describe a focal spot smaller than the wavelength. The benefit of the vectorial representation is that Maxwell's equations are satisfied at any point in space, and analytical expressions for the electric and magnetic field components can be calculated \citep{2010_April, 2015_Salamin, 2015_JeongTM}. Experimental implementation of the tight-focusing scheme by a parabola with f-number, $f_N$, (the ratio of the focal length, $f$, to the beam diameter, $D$) of $0.6$ claims focusing of a $45 \kern0.2em \mathrm{TW}$ laser to a ${\sim} 0.8 \kern0.2em \mathrm{\upmu m}$ focal spot diameter, leading to an intensity of ${\sim} 10^{22} \kern0.2em \mathrm{W cm^{-2}}$ \citep{2004_Bahk}, where a similar intensity is achieved by focusing a $0.3 \kern0.2em \mathrm{PW}$ using a parabola of $f_N = 1.3$ \citep{2017_PirozhkovAS}.

\par Apart from the usually employed linearly polarised (LP) lasers, the radially polarised (RP) and azimuthally polarised (AP) lasers draw much interest of several research groups, employing multi-PW lasers for electron \citep{2010_Salamin, 2012_Payeur} and proton/ion \citep{2010_Salamin, 2012_JianXing, 2015_Ghotra} acceleration. Let us define the laser propagation direction to be along $\bm{\hat{x}}$. In cylindrical coordinates, a RP plane wave satisfies $E_r \bm{\hat{r}} = c B_{\phi} \bm{\hat{\phi}}$ everywhere, where $E_r \bm{\hat{r}}$ is the radial electric field component, $B_{\phi} \bm{\hat{\phi}}$ is the azimuthal magnetic field component and $c$ is the speed of light in vacuum. For the AP laser, the electric and magnetic field components are interchanged. However, under tight-focusing conditions the relation $E_r \bm{\hat{r}} = c B_{\phi} \bm{\hat{\phi}}$ breaks down due to the appearance of a longitudinal electric field component, $E_x \bm{\hat{x}}$ for a RP laser, and a longitudinal magnetic field component, $B_x \bm{\hat{x}}$, for the AP laser \citep{2006_Salamin, 2010b_Salamin, 2018_JeongTM}. Compared to LP lasers, both RP and AP lasers were found experimentally to give a smaller focal spot \citep{2003_Dorn, 2015_Cheng}, in agreement with the elongated electric field distribution for a LP laser \citep{2018_JeongTM}.

\par When the concept of a single-cycle laser is combined with the tight-focusing technique then the $\lambda^3$ regime is obtained, where for a certain laser power one can use minimal energy to achieve the highest intensity \citep{2002_MourouG}. If the $\lambda^3$ regime is applied to an $100 \kern0.2em \mathrm{PW}$ laser, then an intensity exceeding $10^{25} \kern0.2em \mathrm{W cm^{-2}}$ will be achieved. This ultra-intense regime is capable of providing a plethora of particles, such as \textgamma-photons, leptons [electrons ($\mathrm{e^-}$) and positrons ($\mathrm{e^+}$)] and hadrons [protons ($\mathrm{p^+}$) and/or heavy ions ($\mathrm{i^+}$)] \citep{2006_MourouG}. Although \textgamma-photons are achievable even by near-PW class lasers, high laser to \textgamma-photon energy conversion efficiency, $\kappa_\gamma$, is important for applications in photonuclear reactions \citep{2004_NedorezovVG}, astrophysical studies \citep{1992_ReesMJ, 2015_BulanovSV, 2018_PhilippovAA, 2021_AharonianF} and study of the extremely high energy density on materials science \citep{2013_EliassonB}.

\par At laser intensities of ${\sim} \kern0.1em 10^{24} \kern0.2em \mathrm{W cm^{-2}}$ the multiphoton Compton scattering process dominates the \textgamma-photon emission \citep{2013_RidgersCP, 2018_LezhninKV}. During that process, a hot electron/positron is scattered after collision with the incident laser field, its velocity and direction values change and a scattered \textgamma-photon is produced. The process is synopsised in $\mathrm{e^{\pm}} + N \omega_l \rightarrow \mathrm{e^{\pm}} + \omega_\gamma$ where $\omega_l$ is the central laser frequency, $\omega_\gamma$ is the scattered \textgamma-photon frequency and $N >> 1$ is the number of laser photons lost.

\par The Schwinger field represents the field required for the vacuum to break into an $\mathrm{e^- \mbox{-} e^+}$ pair, and it equals $E_S = m_e^2 c^3 / (e \hbar) \approx 1.3 \times 10^{18} \kern0.2em \mathrm{V m^{-1}}$, where $m_e$ is the electron rest mass, $\hbar$ is the reduced Planck constant and $e$ is the elementary charge \citep{1982_BerestetskiiVB}. The probability that a \textgamma-photon will be emitted through multiphoton Compton scattering depends on the parameter \citep{1970_RitusVI}
\begin{equation}
\chi_e =  \sqrt{ \left(\gamma_e \frac{{\bm{E}}}{E_S} + \frac{{\bm{p}}}{m_e} \times \frac{{\bm{B}}}{E_S} \right)^2 - \left(\frac{{\bm{p}}}{m_e c} \cdot \frac{{\bm{E}}}{E_S} \right)^2 } ,
\label{eq:xi_e}
\end{equation}
where $\gamma_e$ is the electron/positron Lorentz factor of momentum $\bm{p}$ prior scattering, ${\bm{B}}$ and ${\bm{E}}$ are the magnetic and electric fields at the position of the electron. For high $\kappa_\gamma$ the condition $\chi_e >> 1$ must be met \citep{2012_NakamuraT, 2012_RidgersCP}. Although the emission model used \citep{2013_RidgersCP} breaks down for ${\alpha \chi_e^{2/3} > 1 }$ \citep{1970_RitusVI, 1979_NarozhnyNB, 2019_IldertonA}, where $\alpha = e^2 / (4 \upi \varepsilon_0 \hbar c)$ is the fine structure constant and $\varepsilon_0$ is the vacuum permittivity, it requires laser intensities significantly higher than those used in the present work.

\par The $\mathrm{e^- \mbox{-} e^+}$ pair generation mechanism in section \ref{ResultsDiscussion} is the multiphoton Breit-Wheeler process \citep{2009_EhlotzkyF}, synopsised in ${ \omega_\gamma + N \omega_l \rightarrow \mathrm{e^{-}} + \mathrm{e^{+}} }$. Here, a large number of laser photons interacts with a high energy \textgamma-photon generated earlier through multiphoton Compton scattering, and then generates an $\mathrm{e^- \mbox{-} e^+}$ pair. The probability of a \textgamma-photon to produce a pair is governed by the parameter \citep{1970_RitusVI}
\begin{equation}
\chi_\gamma =  \frac{\hbar \omega_l}{m_e c^2} \sqrt{ \left( \frac{{\bm{E}}}{E_S} + c {\bm{\hat{p}}} \times \frac{{\bm{B}}}{E_S} \right)^2 - \left({\bm{\hat{p}}} \cdot \frac{\bm{E}}{E_S} \right)^2 } ,
\label{eq:xi_g}
\end{equation}
where $\bm{\hat{p}}$ is the unit vector of the \textgamma-photon momentum.

\par The high fields available by the multi-PW lasers attracted the interest on \textgamma-photon generation. An electron co-propagating with the laser field produces neither \textgamma-photons nor $\mathrm{e^- \mbox{-} e^+}$ pairs due to the opposite contribution of the electric and magnetic terms in equation \eqref{eq:xi_e}. However, in a realistic laser-foil experiment scenario the laser field is reflected on the foil front surface, changing its orientation and therefore enabling generation of \textgamma-photons \citep{2002_ZhidkovA, 2005_KogaJ, 2018_GuYJ}. Another early approach on increasing the \textgamma-photon yield suggested the use of two counter-propagating pulses \citep{2008_BellAR, 2009_KirkJG, 2015_LuoW, 2016_GrismayerT}. This scheme was later generalised in the use of multiple laser beams \citep{2016_VranicM, 2017_GongZ}. The geometry of the target itself was also proven to be crucial as the formation of a preplasma enhanced \textgamma-photon formation \citep{2018_LezhninKV, 2020_WangXB}. Other schemes employing micro-fabrication of the targets taking advantage of the reflected laser field have also been investigated \citep{2019_JiLL, 2021_ZhangLQ}. In addition to the all-optical approach, the combination of a sub-PW laser beam with high-energy electrons is considered \citep{2019_MagnussonJ}.

\par The theoretical framework on the absorption of the energy of a plane wave by the electrons and ions of a foil target is described in reference \citet{1998_VshivkovVA}, although ignoring the energy share to generated \textgamma-photons and consequently the effect of $\mathrm{e^- \mbox{-} e^+}$ pairs. In equation (17) of reference \citet{1998_VshivkovVA}, the target thickness, $l$, is connected to the electron number density, $n_e$, through
\begin{equation}
\epsilon_0 = \frac{\upi n_e l}{n_{cr} \lambda} ,
\label{eq:absorption}
\end{equation}
where $\epsilon_0$ is the normalised areal density and $n_{cr} = \varepsilon_0 m_e \omega^2 /e^2$ is the critical electron number density. The optimum condition for coupling the plane wave to the target is obtained for $\epsilon_0 = a_0$, where $a_0 = e \kern0.1em E / (m_e \kern0.1em c \kern0.1em \omega_l)$ is the dimensionless amplitude. For $\epsilon_0 << a_0$, relativistic transparency of the target results in weak coupling of the laser to the target, whilst for $\epsilon_0 >> a_0$, the laser field is strongly reflected by the target front surface.

\par Equations (32) and (33) in reference \citet{1998_VshivkovVA} give the ratio of the incident (at an angle $\theta_0$ with the target normal) to reflected wave amplitude for an s-polarised laser, $r^s = \varepsilon_0 / [\mathrm{i} \cos(\theta_0)+\varepsilon_0]$, and a p-polarised laser, $r^p = \varepsilon_0 \cos(\theta_0) / [\mathrm{i}+\varepsilon_0 \cos(\theta_0)]$, respectively. In an AP laser, $E_x$ is always zero; in contrary, in a RP laser $E_x$ increases by reducing the f-number. At $\theta_0 = 90 \degree$ there is a qualitative analogy between $r^s$ at $\theta_0 = 0 \degree$ with an AP laser on one hand, and $r^p$ with a RP laser on the other. Therefore, at the tight-focusing scheme, an AP laser is reflected stronger than a RP laser. Up to this point, we have discussed the physical processes enabling us to study the interaction of an ultrarelativistic $\lambda^3$-laser with a solid target via particle-in-cell (PIC) simulations. 

\par One aspect not addressed in PIC simulation studies is the further interactions of the multi-MeV energy particles with the surrounding material, either the vacuum chamber itself or a secondary target. PIC-produced particles generate electrons through ionisation \citep{1944_LandauL} but also $\mathrm{e^- \mbox{-}  e^+}$ pairs through pair production in the Coulomb field of nuclei \citep{1934_BetheH} and/or atomic electrons \citep{1939_WheelerJA}. Post-PIC \textgamma-photons result from either Rayleigh/Compton scattering \citep{1923_ComptonAH} or Bremsstrahlung emission \citep{1959_KochHW,1991_AichelinJ}. Furthermore, neutrons, protons, ions, and  nuclides are produced through photonuclear reactions \citep{1970_HaywardE}, electronuclear reactions \citep{1975_BudnevVM} and through nuclear interactions with heavy ions \citep{1991_AichelinJ}. These interactions are simulated by the Monte Carlo (MC) particle transport code FLUKA \citep{2015_BattistoniG, 2014_BoehlenTT} which can estimate the radioactive nuclides produced and the energy spectra of the post-PIC generated particles. These estimations are useful in nuclear waste management \citep{1998_iaea}, positron annihilation lifetime spectroscopy \citep{2021_AudetTL}, $\mathrm{e^- \mbox{-}  e^+}$ plasma studies \citep{2011_ChenH, 2015_SarriG} and nuclear medicine \citep{2002_SchneiderU}.

\par This paper starts with the description of our numerical solution for the laser field under the tight-focusing scheme as described in references \citet{2015_JeongTM} (for LP lasers) and \citet{2018_JeongTM} (for RP and AP lasers). Based on the choice of a single-cycle pulse, the laser focuses in a ${\sim} \kern0.1em \lambda/2$ diameter sphere ($\lambda^3$ regime), for which an analytical estimation of the peak intensity is obtained. It is found that an ${\sim} \kern0.1em 80 \kern0.2em \mathrm{PW}$ laser leads to a peak intensity of $10^{25} \kern0.2em \mathrm{W cm^{-2}}$. The $\lambda^3$ regime exhibits a complex interaction with the foil target as discussed in section \ref{ElectronEvolution}, regardless the great simplicity of the problem compared to multi-cycle pulses interacting with sophisticated target geometries. Sections \ref{GphotonPositron} and \ref{Gflash} describe the evolution of \textgamma-photon and $\mathrm{e^- \mbox{-} e^+}$ pair generation. Ballistic evolution of the \textgamma-photons reveals a multi-PW \textgamma-flash, expanding with preference to certain directions depending on the laser polarisation mode. A multi-parametric dependency of the laser energy transferred to each particle species is presented in sections \ref{MappingCE}, where the variables include the target thickness, electron number density and laser polarisation. At the optimal parameters combination, $\kappa_\gamma$ is approaching $50 \kern0.2em \%$, accompanied by a laser to positron energy conversion efficiency, $\kappa_{e+}$, of ${\sim} \kern0.1em 10 \kern0.2em \%$. Our results are generalised in section \ref{PowerScaling} for laser powers in the range $1 \kern0.2em \mathrm{PW} \leqslant P \leqslant 300 \kern0.2em \mathrm{PW}$, revealing a saturating trend for $\kappa_\gamma$, along with an optimum region of $\mathrm{e^- \mbox{-} e^+}$ pair avalanche altering the \textgamma-photon spectrum. As a final step, in section \ref{FLUKA} the obtained \textgamma-flash is combined with MC simulations in the vicinity of a high-Z secondary target, to elucidate the importance of the photonuclear interactions.


\section{Simulation Setup} \label{Simulations}


\subsection{Configuration of the $\lambda^3$ Fields} \label{Lambda3}

\par Since the paraxial approximation frequently used by default in PIC codes fails to correctly form the fields in the $\lambda^3$ regime, we followed a method where the electromagnetic fields are pre-calculated based on the tight-focusing scheme. We have obtained numerical solutions to the theory described in reference \citet{2015_JeongTM} for a LP tightly focused laser, where the validity of the model can be applied for $f_N \ge 1/4$. We have then extended our numerical solutions for a RP and AP laser, based on the theoretical solutions in reference \citet{2018_JeongTM}. Here, we describe the basic steps followed in order to calculate the $\lambda^3$-fields on focus, through a Fortran program we developed.

\par We assume a laser before parabola having a uniform spatial profile (a super-Gaussian profile of which the order goes to infinity) of diameter $D$, and that the beam is decomposed to the sum of fundamental wavelengths \cite{2014_BohleF}, corresponding to a minimum wavelength of $\lambda_{min} = 700 \kern0.2em \mathrm{nm}$, a maximum wavelength of $\lambda_{max} = 1750 \kern0.2em \mathrm{nm}$, a central wavelength of $\lambda_{c} = 1000 \kern0.2em \mathrm{nm}$ and equally spaced, equally weighted wavevector intervals (for mathematical simplification) of $dk = (1/\lambda_{min}-1/\lambda_{max})/(\lambda_{max}-\lambda_{min})$.

\par
\begin{figure}
  \centering
  \includegraphics[width=1.00\linewidth]{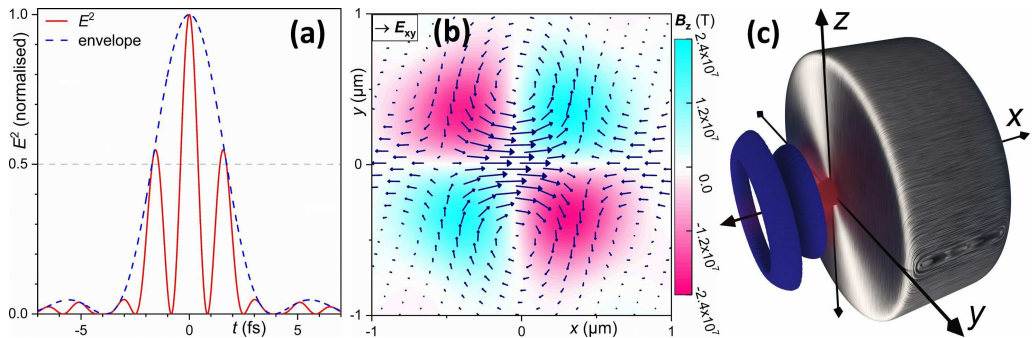}
  \caption{
  (\textit{a}) The $E^2$ profile of the unfocused laser as a function of time is shown by the red line, as described in subsection \ref{Lambda3}. The blue dashed line shows the pulse envelope, with a pulse duration of ${\sim} \kern0.1em 3.4 \kern0.2em \mathrm{fs}$. (\textit{b}) Electromagnetic field representation of the $\lambda^3$-laser, for the laser parameters used in this paper. The black arrows correspond to the electric field vectors, over-plotted on a contour of the magnetic field, on the xy-plane. The result is obtained after free-propagating the externally calculated fields into EPOCH, near focal position (at ${\sim} \kern0.1em -0.3 \kern0.2em \mathrm{fs}$). This field corresponds to a time-averaged peak intensity of $10^{25} \kern0.2em \mathrm{W cm^{-2}}$.  (\textit{c}) Schematic representation of the simulation setup. The grey cylinder represents the target. The blue intensity isosurface at $2 \kern0.1em {\times} \kern0.1em 10^{24} \kern0.2em \mathrm{W cm^{-2}}$ corresponds to the externally imported electric and magnetic fields before propagation. The red intensity isosurface (FWHM of peak intensity) shows the $\lambda^3$-laser, corresponding to fig. \ref{fig:Fields}(b).
  }
\label{fig:Fields}
\end{figure}

 The integral over all wavevectors gives the electric field of the plane wave laser, as
\begin{equation}
E_{pw} (t) = \frac{\sin(2 \upi c t / \lambda_{max})-\sin(2 \upi c t / \lambda_{min})}{t (2 \upi c / \lambda_{max} -2 \upi c / \lambda_{min})} ,
\label{k_int}
\end{equation}
which when squared, corresponds to the intensity as plotted by the red line in figure \ref{fig:Fields}(a). The envelope of the laser is obtained by the Fourier transform of the flat-top spectral power range, resulting  in an electric field envelope of
\begin{equation}
E_{sinc} (t) = \frac{\sin[\upi c t (1/\lambda_{min}-1/\lambda_{max})]}{\upi c t (1/\lambda_{min}-1/\lambda_{max})} ,
\label{k_int}
\end{equation}
while the corresponding intensity is shown by the blue dashed line in figure \ref{fig:Fields}(a) and corresponds to a pulse duration of ${\sim} \kern0.1em 3.4 \kern0.2em \mathrm{fs}$ at full width at half maximum (FWHM).

\par The calculation of electric and magnetic field components is performed in a Cartesian three-dimensional (3D) grid. Let $E_{sum}^2$ be the sum of the squared electric field over all grid locations, for all three Cartesian components. By setting $V$ as the volume of each computational cell, the laser energy corresponding to the electric field is
\begin{equation}
\mathcal{E}_E = \frac{\varepsilon_0 E_{sum}^2}{2} V .
\label{En_E}
\end{equation}
The energy contribution of the magnetic field is equal to that of the electric field, resulting in a laser energy of $\mathcal{E}_l = \varepsilon_0 E_{sum}^2 V$. By knowing the total laser energy, one can weight accordingly each fundamental frequency contribution, with a weight coefficient, $W$. In our specific case, $\mathcal{E}_l = 280 \kern0.2em{J}$, resulting in a laser power of ${\sim} \kern0.1em 80 \kern0.2em \mathrm{PW}$.

\par The core part of our solution is the estimation of the three electric and three magnetic field components, at each cell of a 3D computational grid. To do so,
at each cell we first sum the field contribution from the incident monochromatic electric field on the focusing optic surface over the azimuthal angle ($0 \le \phi < \upi$) and the polar angle ($\theta_{min} \le \theta \le \upi$, where $\theta_{min}$ is given in reference \citet{2015_JeongTM} as a function of $f$ and $D$) and then sum the contribution from each fundamental wavelength. Therefore, a six-fold Do-loop with Open Multi-Processing Application Programming Interface is employed, with the layers order from outer to inner being $y \rightarrow z \rightarrow x \rightarrow \lambda \rightarrow \theta \rightarrow \phi$.

\par Before solving the field integrals, we calculate a set of inter-related quantities independent to the grid position, $κ= 2 \upi / \lambda$, $A=\sin(\theta)/[1-\cos(\theta)]$ and $B=[1-\cos(\theta)]/(2 k f)$. Three simplification variables connected to the grid location are also calculated, $X=\{2 f \cos(\theta) - x [1-\cos(\theta)]\}/(2 f)$, $Y=\{2 f \sin(\theta) \cos(\phi) - y [1-\cos(\theta)]\}/(2 f)$ and $Z=\{2 f \sin(\theta) \cos(\phi) - z [1-\cos(\theta)]\}/(2 f)$. Then, a phase term is calculated, $F= k [x \cos(\theta)+y \sin(\theta)\cos(\phi)+z \sin(\theta)\sin(\phi)]$.

\par The above expressions simplify the integrands (integrated over $\theta$ and $\phi$) from references \citet{2015_JeongTM, 2018_JeongTM} into the form shown in appendix \ref{appA} for a LP laser and in appendix \ref{appB} for a RP laser. For an AP laser we interchange the integrands of the electric and magnetic terms. The electric field of a RP laser along the laser propagation direction is then
\begin{equation}
E_x = \frac{f}{\lambda_c} \sum_{\lambda=\lambda_{min}}^{\lambda_{max}} W \sum_{\phi=0}^{2 \upi} \sum_{\theta=\theta_{min}}^{\upi} I_{Ex-R} ,
\label{intExR}
\end{equation}
(where $I_{Ex-R}$ is given by equation \eqref{REx}) which is scaled by multiplying by $2 \upi (\upi-\theta_{min})/(n_{\theta} n_{\phi})$, where $n_{\theta}$ and $n_{\phi}$ is the number of elements in the $\theta$-array and $\phi$-array, respectively. By calculating $E_x$, $E_y$, $E_z$ in all grid locations we obtain the three arrays containing the components of the electric field, whilst the same process is applied for the magnetic field calculation.

\subsection{Laser Intensity in the $\lambda^3$ Regime} \label{intensity}

\par In order to find an approximate value of the peak laser intensity, $I_p$, we consider only the central peak of the electric field, as shown in figure \ref{fig:Fields}(a) for $-0.8 \kern0.2em \mathrm{fs} \lessapprox t \lessapprox 0.8 \kern0.2em \mathrm{fs}$, containing ${\sim} \kern0.1em (1/3) \kern0.2em \mathcal{E}_l$ at FWHM (temporal profile). In addition, we consider that an Airy function corresponds to ${\sim} \kern0.1em (1/2) \kern0.2em \mathcal{E}_l$ at FWHM (spatial profile). In the $\lambda^3$ regime, the laser field corresponds to a spherical volume, $V_S$, of ${\sim} \kern0.1em \lambda/2$ diameter. The focused fields are obtained by setting $f_N = 1/3$ in section \ref{Lambda3}. By combining the above, and transforming the temporal dimension in spatial, we get
\begin{equation}
I_p = \frac{c \kern0.1em \mathcal{E}_l/6}{V_S} = \frac{8 \kern0.1em c \kern0.1em \mathcal{E}_l}{\upi \lambda^3} .
\label{eq:intensity}
\end{equation}
In this work $\mathcal{E}_l = 280 \kern0.2em \mathrm{J}$ (apart section \ref{PowerScaling}) and $\lambda = 1 \kern0.2em \mathrm{\upmu m}$, where equation \eqref{eq:intensity} gives $I_p {\approx} 2 \kern0.1em {\times} \kern0.1em 10^{25} \kern0.2em \mathrm{W cm^{-2}}$, or a most commonly used time-average intensity (or simply intensity) of $I {\approx} \kern0.1em 10^{25} \kern0.2em \mathrm{W cm^{-2}}$.

\par The peak intensity can also be calculated in the basis of a more strict definition. The spatial boundary of the $\lambda^3$ regime corresponds to the first minima of the Airy function, which  requires reduction to ${\sim} \kern0.1em 83.8 \kern0.2em \% \kern0.2em \mathcal{E}_l$. On the temporal dimension, consideration of only the central peak of the electric field (as previously) requires further reduction to ${\sim} \kern0.1em 44.2 \kern0.2em \% \kern0.2em \mathcal{E}_l$, reducing it to $\mathcal{E}_l \rightarrow 0.838 \kern0.1em {\times} \kern0.1em 0.442 \kern0.1em {\times} \kern0.1em 280 \kern0.2em \mathrm{J} \approx 104 \kern0.2em \mathrm{J}$.

\par The energy fraction contained in the sphere of Gaussian profile in all directions and of radius $r$ and standard deviation $\sigma = \sqrt{8 \ln(2)} \kern0.2em \mathrm{FWHM}$ can be calculated as:
\begin{multline}
\int_0^{2 \upi} \int_0^{\upi} \int_0^r \left (\sigma \sqrt{2 \upi} \right)^{-3} \exp \left[-\frac{1}{2} \left(\frac{r}{\sigma}\right)^2 \right] r^2 \sin(\theta) \,dr\,d\theta\,d\phi =
\\
\erf \left( \frac{r}{\sqrt{2} \sigma} \right) - \sqrt{\frac{2}{\upi}} \frac{r}{\sigma} \exp \left[-\frac{1}{2} \left( \frac{r}{\sigma} \right)^2 \right] ,
\label{eq:integral}
\end{multline}
By dividing equation \eqref{eq:integral} by the volume of the sphere, taking the limit as $r \rightarrow 0$, and using L'Hospital's rule once, we estimate
\begin{equation}
\lim_{r \rightarrow 0} \frac{\erf \left( \frac{r}{\sqrt{2} \sigma} \right) - \sqrt{\frac{2}{\upi}} \frac{r}{\sigma} \exp \left[-\frac{1}{2} \left( \frac{r}{\sigma} \right)^2 \right]}{ \frac{4}{3} \upi r^3 } =
\frac{1}{(2 \upi \sigma^2)^{3/2}} .
\label{eq:limit}
\end{equation}
By considering the energy contained in the sphere, and transforming the spatial dimension in temporal, we obtain:
\begin{equation}
I_p = \frac{c \kern0.1em (0.838 \kern0.1em {\times} \kern0.1em 0.442 \kern0.1em {\times} \kern0.1em \mathcal{E}_l)}{(2 \upi \sigma^2)^{3/2}} .
\label{eq:intensity2}
\end{equation}
By replacing $\sigma \approx \lambda / [4 \sqrt{2 \ln(2)}]$, equation \eqref{eq:intensity2} gives:
\begin{equation}
I_p = \left[ \sqrt{\frac{\ln(2)}{\upi}} \frac{4}{\lambda} \right]^3 c \kern0.1em (0.838 \kern0.1em {\times} \kern0.1em 0.442 \kern0.1em {\times} \kern0.1em \mathcal{E}_l) \approx \frac{2.457 \kern0.1em c \kern0.1em \mathcal{E}_l}{\lambda^3},
\label{eq:intensity3}
\end{equation}
which again gives $I {\approx} \kern0.1em 10^{25} \kern0.2em \mathrm{W cm^{-2}}$.

\par By relating the intensity to the corresponding electric field through $E = \sqrt{2 \kern0.1em I / (c \kern0.1em \varepsilon_0)}$, the focused laser gives $E {\approx} \kern0.1em 8.7 \kern0.1em {\times} \kern0.1em 10^{15} \kern0.2em \mathrm{V m^{-1}}$. This field gives a value for the dimensionless amplitude of $a_0 \approx 2700$, where we further approximate $\gamma_e \approx a_0$.


\subsection{PIC Simulation Setup} \label{Setup}

\par The results presented in this paper are obtained through 3D PIC simulations by use of the EPOCH \citep{2015_ArberTD} code. The code is compiled with the flags for Quantum Electrodynamics (QED) \citep{2014_RidgersCP} and Higuera-Cary (HC) \citep{2017_HigueraAV} preprocessor directives enabled. The QED module enables \textgamma-photon and $\mathrm{e^- \mbox{-} e^+}$ pair generation, the inclusion of which is essential at ultra-high intensities. Since \textgamma-photon generation is directly connected with electron/positron energy and trajectory, an accurate estimation of their motion is necessary. The HC solver accounts for the necessity of increased motion accuracy, since the default Boris \citep{Boris1970} solver is less reliable for relativistic particles.

\par No laser-block is used in our simulations. Instead, we take advantage of the EPOCH fields-block, which enables the import of a desired electromagnetic field configuration as three electric and three magnetic field components. The field data were pre-calculated (as described in section \ref{Lambda3}) in a 3D grid matching the number of cells per dimension with those used in the PIC grid. In this work we define that the laser is focused at $t=0 \kern0.2em \mathrm{fs}$, as shown in figure \ref{fig:Fields}(b). The imported unfocused field data were calculated at $t \approx -4.27 \kern0.2em \mathrm{fs}$. The simulation setup shown in figure \ref{fig:Fields}(c), where the imported fields are overlapped to the target geometry.

\par The 3D EPOCH grid is cubic, with the focal spot defined at the centre of the cube. All three dimensions extend from $-5.12 \kern0.2em \mathrm{\upmu m}$ to $5.12 \kern0.2em \mathrm{\upmu m}$ with 1024 cells per dimension. The resulting cells are cubes with an edge of $\alpha_c = 10 \kern0.2em \mathrm{nm}$. The highest electron number density used is $5 \times 10^{24} \kern0.2em \mathrm{cm^{-3}}$, for which, at an intensity of $10^{25} \kern0.2em \mathrm{W cm^{-2}}$, the relativistically corrected skin depth is resolved with an accuracy of more than 10 cells per skin depth. At that electron number density, the skin depth can be resolved even with intensities as low as $10^{21} \kern0.2em \mathrm{W cm^{-2}}$. The simulation stops after $16 \kern0.2em \mathrm{fs}$, since beyond that time fields start escaping the simulation box, for which we have set open boundary conditions. The box dimensions are chosen large enough that the laser to each particle species energy conversion efficiency, $\kappa$, saturates.

\par The particle species set at code initialisation are ions and electrons, while \textgamma-photons and $\mathrm{e^- \mbox{-} e^+}$ pairs are generated during code execution. The ion atomic number is set to $Z=1$, while its mass number at $A=2.2$, which is the average $A/Z$ for solid elements with $Z< 50$. EPOCH behaviour was tested for multiplying Z and A by a factor and simultaneously reducing the ion number density by the same factor, giving identical results. Therefore, our simulations can be generalised for most target materials used in laser-matter interaction experiments.

\par The target geometry is cylindrical, with the cylinder radius being $r=2.4 \kern0.2em \mathrm{\upmu m}$ and the height of the cylinder (target thickness), $l$, varying in the range $ 0.2 \kern0.2em \mathrm{\upmu m} \le l \le 2 \kern0.2em \mathrm{\upmu m}$. Although the target can be considered as mass-limited, its radius is large enough that its periphery survives the laser-foil interaction by the end of the simulation. The target front surface is placed at $x=0 \kern0.2em \mathrm{\upmu m}$, coinciding with the focal spot. The electron number density is uniform for each simulation, and is within the range $2 \times 10^{23} \kern0.2em \mathrm{cm^{-3}} \le n_e \le  5 \times 10^{24} \kern0.2em \mathrm{cm^{-3}}$. In order to have 8 macroparticles per cell, the number of ions and initial electrons is set to $8 \upi r^2 l / \alpha_c$. Since it was found that \textgamma-photons with energy $< 1 \kern0.2em \mathrm{MeV}$ account for ${\sim} \kern0.1em 1 \kern0.2em \%$ of the \textgamma-photon energy, only those above that energy threshold were allowed in the simulation


\section{Results and Discussion} \label{ResultsDiscussion}


\par The present section provides a detailed description on the interaction of the ultra-intense laser with a solid target in the $\lambda^3$ regime, for RP, LP and AP lasers. In subsections \ref{ElectronEvolution}, \ref{GphotonPositron} and \ref{Gflash} the description is made for a relatively thick target ($2 \kern0.2em \mathrm{\upmu m}$) with an electron number density similar to titanium ($1.2 \kern0.1em \times \kern0.1em 10^{24} \kern0.2em \mathrm{cm^{-3}}$).


\subsection{Electron Evolution} \label{ElectronEvolution}

\par A schematic representation of the simulation setup used in the current subsection is shown in figure \ref{fig:Fields}(c), where a $\lambda^3$-pulse interacts with a $2 \kern0.2em \mathrm{\upmu m}$ thick cylindrical target of $1.2 \times 10^{24} \kern0.2em \mathrm{cm^{-3}}$ electron number density. These target parameters correspond to the highest $\kappa_\gamma$ achieved in our simulations for an ${\sim} \kern0.1em 80 \kern0.2em \mathrm{PW}$ laser, approaching $50 \kern0.2em \%$. The interaction results in a double exponentially decaying electron spectrum for all three polarisations, where the first exponential is approximately in the energy range of $200 \kern0.2em \mathrm{MeV} \leqslant \mathcal{E}_e \leqslant 500 \kern0.2em \mathrm{MeV}$ and the second for $\gtrapprox 500 \kern0.2em \mathrm{MeV}$. The temperature of the lower energy part of the spectrum is ${\sim} \kern0.1em 100 \kern0.2em \mathrm{MeV}$ and approximately double for the higher energy part. These electrons are accompanied by an ion spectrum of similar temperature, a Maxwell-Juttner-like positron spectrum and a \textgamma-photon exponentially decaying spectrum of ${\sim} \kern0.1em 150 \kern0.2em \mathrm{MeV}$ temperature. The exact temperatures for electron and \textgamma-photon spectra for RP, LP and AP lasers are summarised in table \ref{tab:Temperatures}.

\par As mentioned earlier in section \ref{Introduction}, one fundamental difference of a RP and an AP laser (tightly focused) is the presence and the absence of $E_x$, respectively \citep{2018_JeongTM}. For a LP laser of the same power, although resulting in higher intensity, $E_x$ is weaker than that of the RP laser. For a tightly focused laser, $E_x$ dominates over $E_r$, as seen by the centre of figure \ref{fig:Fields}(b). Another field feature for the tight-focusing scheme is the curled field vectors centred at a distance of ${\sim} \kern0.1em \lambda/2$ from focus. This pattern can be realised as an interference of the Airy pattern for a plane wave, when tightly focused. For the AP laser, the electric and magnetic field roles are interchanged, where the electric field now has a rotating form around the laser propagation axis.

\par Figure \ref{fig:Fields}(b) reveals the complexity of the $\lambda^3$-laser due to interplay of all three field components, versus two for weak-focusing. Furthermore, the single-cycle condition breaks the repetitive nature of a multi-cycle laser, where despite limiting the laser-foil interaction in the wavelength timescale, each time has a unique effect on the evolution of the interaction. That complicated field behaviour results in a significantly different laser-foil interaction, depending on the laser polarisation. For RP, LP and AP lasers $\kappa$ is significantly different, since the electron trajectories are completely incomparable.

\par Let us consider the case of a RP laser. As a result of the laser-foil interaction a conical-like channel is progressively drilled on the foil target by the laser field, where the ejected electrons are either rearranged in the form of a low density pre-plasma distribution, or reshaped as thin over-dense electron fronts. The conical channel formation is mainly mandated by $E_x$, although its formation initiates by the pulse edges even prior the arrival of the focused pulse. The dimensions of the channel are in agreement with the pulse extent, of ${\sim} \kern0.1em \lambda/2$.

\begin{table}
  \begin{center}
\def~{\hphantom{0}}
  \begin{tabular}{lccc}
      {}  & $e^{{}-{}}$ ($\mathcal{E}_e < 500 \kern0.2em \mathrm{MeV}$)   &   $e^{{}-{}}$ ($ \mathcal{E}_e >500 \kern0.2em \mathrm{MeV}$) & \textgamma-photon ($ \mathcal{E}_\gamma >500 \kern0.2em \mathrm{MeV}$) \\[3pt]
       RP laser        &  ~$99 \kern0.2em \mathrm{MeV}$ & ~~$204 \kern0.2em \mathrm{MeV}$ ~ & $139 \kern0.2em \mathrm{MeV}$ \\
       LP laser        &  $128 \kern0.2em \mathrm{MeV}$ & ~~$172 \kern0.2em \mathrm{MeV}$ ~ & $178 \kern0.2em \mathrm{MeV}$ \\
       AP laser       &  ~$96 \kern0.2em \mathrm{MeV}$ & ~~$280 \kern0.2em \mathrm{MeV}$ ~ & $147 \kern0.2em \mathrm{MeV}$ \\
  \end{tabular}
  \caption{The temperature of electrons and \textgamma-photons for a RP, a LP and an AP laser.}
  \label{tab:Temperatures}
  \end{center}
\end{table}

\par The channel formation is considered in three time intervals of $t_a < - \lambda /(4c)$, $-\lambda /(4c) \leqslant t_b \leqslant \lambda /(4c)$ and $t_c > \lambda /(4c)$. At $t_a$, although the peak laser field has not yet reached the focal spot, a low amplitude electric field exist due to the $\sinc$ temporal profile (see figure \ref{fig:Fields}(a)). Those pulses, although several orders of magnitude lower than the peak laser field, are still capable of heating and driving electrons out of the target. In addition, the field corresponding to the outer Airy disks of the main pulse is also capable of affecting the target electrons. Their combined effect is deformation of the steep flat target density profile. At $-1.3 \kern0.2em \mathrm{fs}$ the target profile consists of a sub-micron under-dense region at the target front surface, followed by an over-dense 10s of nanometres thick electron pile-up and then by the rest of the intact target. At that stage a directional ring of high energy electrons also appears at ${\sim} \kern0.1em 60 \degree$ to the target normal, connected with the focusing conditions ($f_N = 1/3$) of the laser field. Finally, a high energy electron population is moving along the laser propagation axis. The momentum of all electron groups is governed by a characteristic time interval of $\lambda /(4c)$.

\par
\begin{figure}
  \centering
  \includegraphics[width=1.00\linewidth]{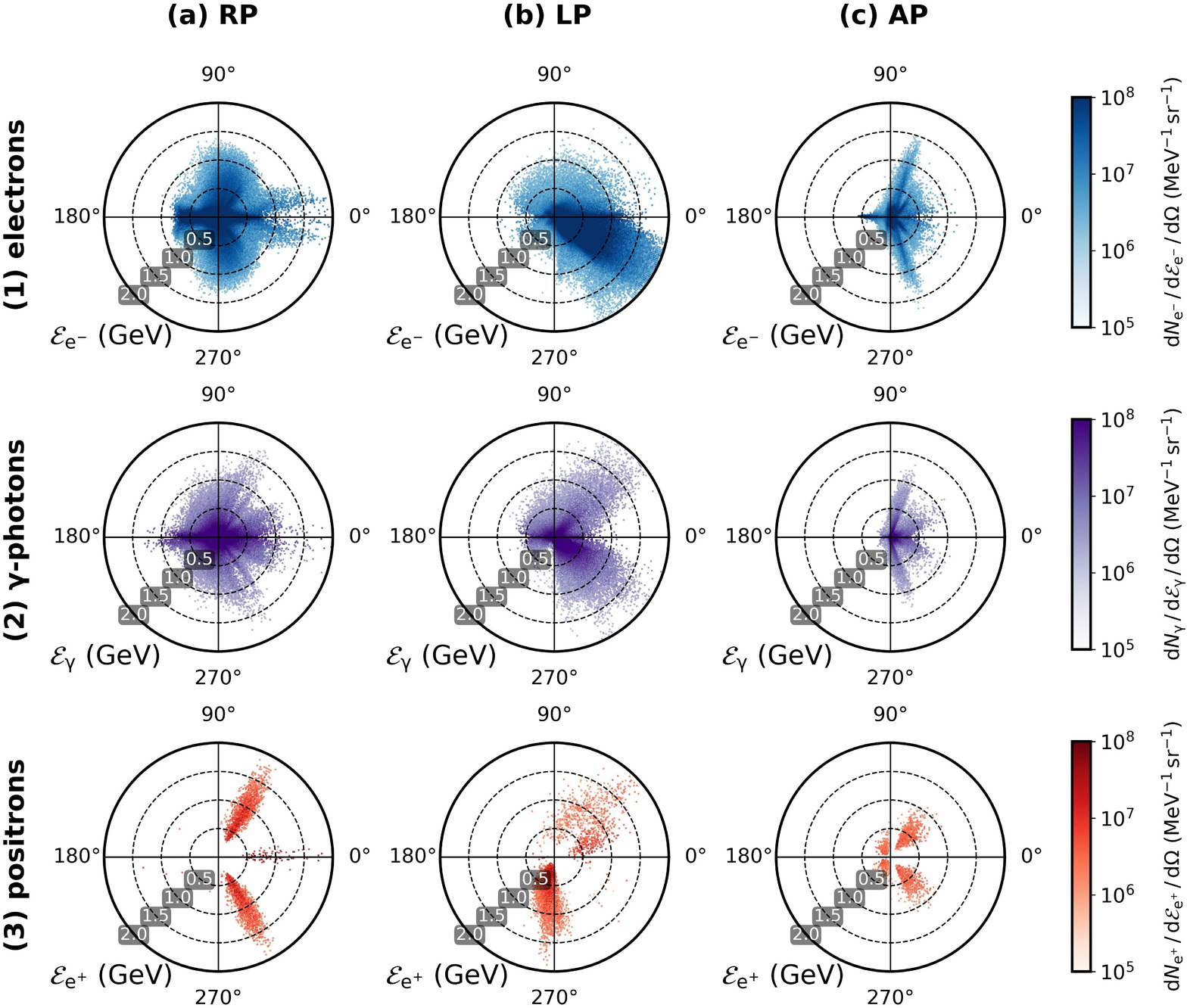}
  \caption{
  Polar energy spectrum diagrams of (\textit{1}) electrons at ${\sim} \kern0.1em 0.7 \kern0.2em \mathrm{fs}$. Polar energy spectrum diagrams of (\textit{2}) \textgamma-photons and (\textit{3}) positrons generated in the time interval $-0.3 \kern0.2em \mathrm{fs} \leqslant t \leqslant 0.7 \kern0.2em \mathrm{fs}$. The figure columns correspond to (\textit{a}) a RP laser, (\textit{b}) a LP laser and (\textit{c}) an AP laser. Animation for a larger time interval is provided in movie-1.
  }
\label{fig:Electrons}
\end{figure}

 The upper row of figure \ref{fig:Electrons} shows the polar energy spectrum of electrons for three polarisations at $0.7 \kern0.2em \mathrm{fs}$. At $t_b$, the curled part of the electric field changes the directionality and distribution of the thin electron ring population, transforming it into a toroidal-like electron distribution with a torus radius of ${\sim} \kern0.1em \lambda/2$, matching the centre of the curled field. Simultaneously, the peak $E_x$ reaches the focal spot without any significant decay, since the toroidal-like electron distribution allows for a practically vacuum region for the field to propagate at. At $-0.3 \kern0.2em \mathrm{fs}$ the electron energy distribution reaches energies of ${\sim} \kern0.1em 1 \kern0.2em \mathrm{GeV}$. However, after a time of $\lambda / (4 c)$ the pulse is reflected by the thin over-dense electron front. By the time the pulse is reflected, the electron population corresponding to the toroidal structure emerges into a closed high energy electron distribution, which can be considered as a pre-plasma at the target front surface.

\par Within $t_b$, high amplitude oscillations of the electron momentum occur. At $t_c$ electron momentum oscillations become gradually less significant, with the electron spectrum eventually saturating. At this stage, the peak laser field is not completely reflected, but $E_x$ starts forming a cavity beyond the over-dense electron front. Part of the laser field then reaches within the cavity, further expanding it. The initial times of this process witness instantaneous intensities an order of magnitude higher than the intensity expected on focus, due to interference of the laser fields after diffraction/reflection by the cavity walls. Although the intensity occurs only instantaneously, it was found to be ${\sim} \kern0.1em 8.8  \kern0.1em \times \kern0.1em 10^{25} \kern0.2em \mathrm{W cm^{-2}}$ in a region approximated by a sphere of ${\sim} \kern0.1em 50 \kern0.2em \mathrm{nm}$ diameter, at $1.7 \kern0.2em \mathrm{fs}$. At this stage, another electron population emerges, driven by the reflected field in the backward direction. In summary, during all stages of the laser-target interaction, electron populations at $0 \degree, \kern0.1em {\sim} \kern0.1em 60 \degree$, and $180 \degree$ are recorded.

\par So far, we have given a detailed explanation of the electron evolution under the influence of a RP $\lambda^3$-laser. For a LP $\lambda^3$-laser, although the $E_x$ still does exist, the lack of rotational symmetry does not allow the curled fields to take a toroidal form. Therefore, although a pre-plasma distribution is formed, it is extremely asymmetric along the laser oscillation direction. The thin over-dense electron pile-up is also asymmetric. The asymmetry is due to the initial decay of the flat target, diverting the laser into a favourable direction. Asymmetric field interference does not allow the laser to form a conical cavity, but the random nature of the process forms a macroscopically rectangle-like cavity instead.

\par For the case of an AP $\lambda^3$-laser the cavity formation is simpler. The absence of $E_x$ means that the laser can be absorbed by the target in a similar manner to a weakly focused laser, suppressing the target deformation. The deformation takes the form of an over-dense electron pile-up without pre-plasma. The pre-plasma created is also suppressed, in a region near the laser propagation axis. However, by the end of the simulation a cavity is eventually created, although by that time strong fields do not exist and $\kappa_\gamma$ is limited, as seen in section \ref{GphotonPositron}.

\subsection{\textgamma-photon and Positron Evolution} \label{GphotonPositron}

\par At ultra-high laser intensities \textgamma-photon and $\mathrm{e^- \mbox{-} e^+}$ pair generation plays an important role in the laser-target interaction. The non-trivial form of the $\lambda^3$-field reveals a strong dependency of \textgamma-photon and $\mathrm{e^- \mbox{-} e^+}$ pair generation every quarter-period, in connection with the altered gradient/sign of the laser field, which in extension defines the electron motion as seen in section \ref{ElectronEvolution}.

\par The \textgamma-photon generation can be visualised by a series of polar energy spectra diagrams. An animation for various times is provided as supplementary material in movie-1. However, since we are mainly interested in the evolution of \textgamma-photon generation, it is more appropriate to consider the difference of every two subsequent polar diagrams, where our simulations output the data every $1 \kern0.2em \mathrm{fs}$, a time-interval similar to the quarter-period of $5/6 \kern0.2em \mathrm{fs}$. The second row of figure \ref{fig:Electrons} (see also the second row in movie-1) shows these diagrams for the three polarisations used, for a time interval of $-0.3 \kern0.2em \mathrm{fs} \leqslant t \leqslant 0.7 \kern0.2em \mathrm{fs}$. These diagrams have the benefit of not only showing at which angle \textgamma-photons are generated at, but also a negative value exhibits \textgamma-photon loss. In our simulations no \textgamma-photons are allowed to escape the simulation and lack of a \textgamma-photons is attributed only to an $\mathrm{e^- \mbox{-} e^+}$ pair formation. The corresponding plots for positrons are shown in the third row of figure \ref{fig:Electrons}. We must clarify that \textgamma-photons and $\mathrm{e^- \mbox{-} e^+}$ pairs are not only formed in positive and negative polar diagram values, respectively, but a negative sign means that more \textgamma-photons are lost to $\mathrm{e^- \mbox{-} e^+}$ pairs than what generated by the multiphoton Compton scattering process.

\par Let us consider a RP laser. Initially, up-to $-2.3 \kern0.2em \mathrm{fs}$, only a small fraction of electrons obtains relativistic energies due to the low amplitude periphery of the $\lambda^3$ field. These electrons then interact with the reflected relatively low amplitude edge of the laser \citep{2012_RidgersCP} producing low energy (${\sim} \kern0.1em 0.1 \kern0.2em \mathrm{GeV}$) \textgamma-photons. However, at the next femtosecond significantly more electrons acquire relativistic energies and in combination with the increased amplitude of the field as approaching the focal spot at ${\sim} \kern0.1em 60 \degree$, directional \textgamma-photons of ${\sim} \kern0.1em 0.5 \kern0.2em \mathrm{GeV}$ appear at the same angle. In addition, another energetic electron population appears towards the laser propagation axis, producing another high energy \textgamma-photon population.

\par The similar process continues up-to $-0.3 \kern0.2em \mathrm{fs}$, although electric fields are intensified giving \textgamma-photons of ${\sim} \kern0.1em 1 \kern0.2em \mathrm{GeV}$. The newly generated \textgamma-photons are still oriented purely at a ${\sim} \kern0.1em 60 \degree$ cone and also on the laser axis. It is no surprise that the \textgamma-photon yield continues increasing until the laser pulse peak amplitude reaches the focal spot. What is of a surprise is that the high energy part of the \textgamma-photon spectrum drops near. The overall increase in $\kappa_\gamma$ is mostly due to an isotropic generation of the moderate to low energy \textgamma-photons.

\par In figure \ref{fig:Electrons}(c1-c3), one can observe the polar energy spectrum of positrons generated within $1 \kern0.2em \mathrm{fs}$ time interval at ${\sim} \kern0.1em 60 \degree$, corresponding to the conversion of high energy \textgamma-photons to $\mathrm{e^- \mbox{-} e^+}$ pairs. Strong $\mathrm{e^- \mbox{-} e^+}$ pair generation continues within the next two femtoseconds and then sharply decreases. This time interval is characterised by a region of negative values (\textgamma-photon loss) in the high energy part of the \textgamma-photon energy spectra produced within a finite time, when plotted as a function of time. This plot (not shown) reveals the quarter-period behaviour of \textgamma-photon generation as a superposition of several peaks. As the field amplitude drops, the \textgamma-photon production rate also drops. One can approximate the \textgamma-photon production rate as a steep Gaussian-like function up-to the focus, followed by an exponentially-like decay.

\par As mentioned in section \ref{Introduction}, a LP laser results in a higher peak intensity compared to a RP laser of the same power. Although the lack of symmetry results in a weaker coupling of the laser energy to the target electrons, the higher intensity on focus results in a slight enhancement of the high energy part of the \textgamma-photon energy spectrum for the LP laser case. However, at energies lower than ${\sim} \kern0.1em 0.37 \kern0.2em \mathrm{GeV}$ the amplitude of the \textgamma-photon energy spectrum is higher for the RP laser case. Consider that for the RP laser, \textgamma-photons with energy ${< \sim} \kern0.1em 0.37 \kern0.2em \mathrm{GeV}$ contain ${\sim} \kern0.1em 90 \kern0.2em \%$ of the \textgamma-photon energy. Therefore, although the LP laser results in higher cut-off energies, it results in $\kappa_\gamma$ of ${\sim} \kern0.1em 40 \kern0.2em \%$, compared to ${\sim} \kern0.1em 47 \kern0.2em \%$ for a RP laser. For the AP laser, although strong fields do exist, the Lorentz factor of electrons is significantly lower than the other two polarisation cases. Furthermore, no significant pre-plasma is formed in the laser field reflection region. As a result, only a $\kappa_\gamma$ of ${\sim} \kern0.1em 20 \kern0.2em \%$ occurs.

\par The positron spectra for LP and RP lasers overlap, apart in the very high and very low parts of the spectra, where the positrons obtain ${\sim} \kern0.1em 7 \kern0.2em \%$ and ${\sim} \kern0.1em 9 \kern0.2em \%$ $\kappa_{e+}$ by the end of the simulation. However, this energy is not purely a result of \textgamma-photon energy conversion to $\mathrm{e^- \mbox{-} e^+}$ pairs, but it is also a result of acceleration/deceleration of those positrons by the laser field, in the same manner as electrons \citep{2012_RidgersCP}. One index that can directly compare two interactions is the number of positrons generated, regardless their energy, where for a RP laser and a LP laser we obtain ${\sim} \kern0.1em 5.7 \kern0.1em {\times} \kern0.1em 10^{11}$ positrons and ${\sim} \kern0.1em 4 \kern0.1em {\times} \kern0.1em 10^{11}$ positrons, respectively. In comparison, the AP laser results in ${\sim} \kern0.1em 3 \kern0.2em \%$ $\kappa_{e+}$, but generation of only ${\sim} \kern0.1em 1.9 \kern0.1em {\times} \kern0.1em 10^{11}$ positrons. The imbalance of $\kappa_{e+}$ to their number for the various laser polarisation modes verifies that positrons are strongly affected by the laser field after their generation.

\par In addition, our simulations record local positron number densities as high as ${\sim} \kern0.1em 3 \kern0.1em {\times} \kern0.1em 10^{26} \kern0.2em \mathrm{cm^{-3}}$, approximately two orders of magnitude higher than the titanium target electron number density, emphasizing the collective effect of $\mathrm{e^- \mbox{-} e^+}$ pairs in the laser-target interaction. By assuming that the $\mathrm{e^- \mbox{-} e^+}$ pairs are contained in a uniform density sphere of diametre equal that of the $\lambda^3$-laser, they correspond to an average number density of ${\sim} \kern0.1em 10^{25} \kern0.2em \mathrm{cm^{-3}}$, still an order of magnitude higher than the target electron number density.


\subsection{\textgamma-flash} \label{Gflash}

\par As mentioned in reference \citet{2021_HadjisolomouP}, the \textgamma-photons generated during the interaction of a RP $\lambda^3$-laser with a foil appear in the form of a spherically expanding shell. The \textgamma-photon energy density of this shell is not uniform since more energetic \textgamma-photons are at $0 \degree, 180 \degree, {\sim} \kern0.1em 60 \degree$. Computational constrains limit the \textgamma-photon shell expansion within a cube of $\pm 5.12 \kern0.2em \mathrm{\upmu m}$ edges. In EPOCH code, if a \textgamma-photon is not lost to an $\mathrm{e^- \mbox{-} e^+}$ pair, then it propagates ballistically. Therefore, the \textgamma-photon located at position $(x_{i,1},y_{i,1},z_{i,1})$ can propagate a distance, $\mathcal{D}$, to a new position, $(x_{i,2},y_{i,2},z_{i,2})$ (where the subscript $i$ denotes the corresponding \textgamma-photon of energy $\mathcal{E}_i$), as
\begin{equation}
x_{i,2} = x_{i,1} + \mathcal{D} p_{i,x} / \sqrt{p_{i,x}^2+p_{i,y}^2+p_{i,z}^2} , 
\end{equation}
\begin{equation}
y_{i,2} = y_{i,1} + \mathcal{D} p_{i,y} / \sqrt{p_{i,x}^2+p_{i,y}^2+p_{i,z}^2} , 
\end{equation}
\begin{equation}
z_{i,2} = z_{i,1} + \mathcal{D} p_{i,z} / \sqrt{p_{i,x}^2+p_{i,y}^2+p_{i,z}^2} ,
\end{equation}
which corresponds to a new distance, $r_i$, from the axis origin, 

\begin{figure}
  \centering
  \includegraphics[width=1.00\linewidth]{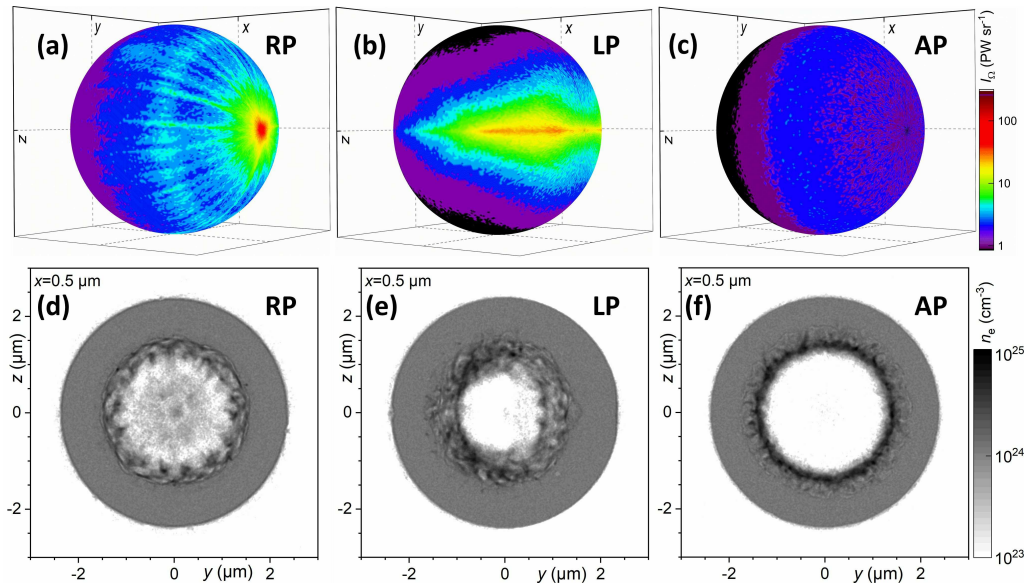}
  \caption{
  \textgamma-photon radiant intensity for (\textit{a}) a RP laser, (\textit{b}) a LP laser and (\textit{c}) an AP laser, $64 \kern0.2em \mathrm{fs}$ after the start of the simulation. Electron number density cross-section at $x = 0.5 \kern0.2em \mathrm{\upmu m}$, for (\textit{d}) a RP laser, (\textit{e}) a LP laser and (\textit{f}) an AP laser, at the end of the simulation.
  }
\label{fig:Flash}
\end{figure}

\par The ballistic \textgamma-photon expansion for a $\mathcal{D}$ of $15.36 \kern0.2em \mathrm{\upmu m}$ reveals that the \textgamma-photon population at $60 \degree$ rapidly decreases geometrically. However, the \textgamma-photon populations at $0 \degree, 180 \degree$ due to their small solid angle are preserved, as shown in figure \ref{fig:Flash}(a). The spherically expanding \textgamma-flash at large distances is considered as originating from a virtual point source, although as seen in section \ref{GphotonPositron} the \textgamma-photons are not generated instantaneously.

\par As mentioned earlier for a RP laser, \textgamma-photons obtain ${\sim} \kern0.1em 47 \kern0.2em \%$ of the ${\sim} \kern0.1em 280 \kern0.2em \mathrm{J}$ laser energy, or in other words, the \textgamma-flash energy is ${\sim} \kern0.1em 130 \kern0.2em \mathrm{J}$. To calculate the mean location of the \textgamma-flash, $\mu$, we calculate the first order moment as
\begin{equation}
 \mu = \frac{\sum\limits_i \mathcal{E}_i r_i}{\sum\limits_i \mathcal{E}_i} ,
\end{equation}
which for the RP laser case gives $\mu = {\sim} \kern0.1em 18.5 \kern0.2em \mathrm{\upmu m}$.

\par The second order moment gives the position variance, $\sigma^2$, of the \textgamma-flash, as
\begin{equation}
 \sigma^2 = \frac{\sum\limits_i \mathcal{E}_i (r_i -\mu)^2}{\sum\limits_i \mathcal{E}_i} ,
\end{equation}
while the square root of the variance gives the standard deviation, which in turn gives the temporal FWHM of the \textgamma-flash. For a RP laser the \textgamma-flash has a FWHM duration of ${\sim} \kern0.1em 4.2 \kern0.2em \mathrm{fs}$ resulting in a ${\sim} \kern0.1em 31 \kern0.2em \mathrm{PW}$ \textgamma-flash.

\begin{table}
  \begin{center}
\def~{\hphantom{0}}
  \begin{tabular}{lccccc}
      {} & $\mathcal{E}_{\gamma}$ &  $\mu$ & $\sigma$ & $t_{FWHM}$ & $P$ \\[3pt]
       RP laser        & ~$131 \kern0.2em \mathrm{J}$ & ~$18.6 \kern0.2em \mathrm{\upmu m}$ & $0.53 \kern0.2em \mathrm{\upmu m}$ & $4.2 \kern0.2em \mathrm{fs}$ & $31 \kern0.2em \mathrm{PW}$ \\
       LP laser        & ~$113 \kern0.2em \mathrm{J}$ & ~$18.6 \kern0.2em \mathrm{\upmu m}$ & $0.52 \kern0.2em \mathrm{\upmu m}$ & $4.1 \kern0.2em \mathrm{fs}$ & $28 \kern0.2em \mathrm{PW}$ \\
       AP laser   & ~~$58 \kern0.2em \mathrm{J}$ & ~$18.4 \kern0.2em \mathrm{\upmu m}$ & $0.58 \kern0.2em \mathrm{\upmu m}$ & $4.5 \kern0.2em \mathrm{fs}$ & $13 \kern0.2em \mathrm{PW}$ \\
  \end{tabular}
  \caption{Energy, mean position, position variance, duration and power of the \textgamma-flash for a RP, a LP and an AP laser.}
  \label{tab:moments}
  \end{center}
\end{table}

\par For a LP laser and an AP laser the \textgamma-flash power is ${\sim} \kern0.1em 28 \kern0.2em \mathrm{PW}$ and ${\sim} \kern0.1em 13 \kern0.2em \mathrm{PW}$, respectively. The AP laser results in high energy \textgamma-photons emitted mainly at ${\sim} \kern0.1em 60 \degree$, while the dominant low energy \textgamma-photons are emitted isotropically, as shown in figure \ref{fig:Flash}(c). The LP laser case results in two detached \textgamma-photon fronts delayed by half-period, at ${\sim} \kern0.1em \pm 45 \degree$ and with higher \textgamma-photon energy density on the plane defined by the laser field oscillation. At large distances, these fronts merge, and therefore, expand as thin rings, as seen in figure \ref{fig:Flash}(b). The energy, mean position, position variance, duration and power of the \textgamma-flash for a RP, a LP and an AP laser are summarised in table \ref{tab:moments}.

\par The electron number density for the RP laser case forms radially symmetric regular modulations inside the target cavity [figure \ref{fig:Flash}(d)]. The effect of those modulations is reflected in the \textgamma-photon radiant intensity distribution, as shown in figure \ref{fig:Flash}(a). For the LP laser case, although electron modulations are formed, they are symmetric only with respect to the laser oscillation direction [figure \ref{fig:Flash}(e)]. Therefore, radial \textgamma-photon modulations are not observed [figure \ref{fig:Flash}(b)] and any \textgamma-photon modulation is hidden by the macroscopic \textgamma-photon distribution. For the AP laser case, radial electron modulations are formed, but with outwards directionality. Furthermore, they are shielded by the field region by an overdense electron ring distribution [figure \ref{fig:Flash}(f)]. As a result, no obvious \textgamma-photon modulations are observed.


\subsection{Mapping the Energy Conversion Efficiency} \label{MappingCE}

\par In the current subsection we present the results of our multi-parametric study for an ${\sim} \kern0.1em 80 \kern0.2em \mathrm{PW}$ laser (RP, LP and AP laser cases) on $\kappa_\gamma$, $\kappa_{e+}$, laser to electron energy conversion efficiency, $\kappa_{e-}$ and laser to ion energy conversion efficiency, $\kappa_{i+}$. The variable parameters include the target thickness and electron number density, for which the inversely proportional relation is mentioned in section \ref{Introduction}. The results are presented in the form of ternary plots \citep{1982_WestD} accompanied by radar charts.

\par
\begin{figure}
  \centering
  \includegraphics[width=0.94\linewidth]{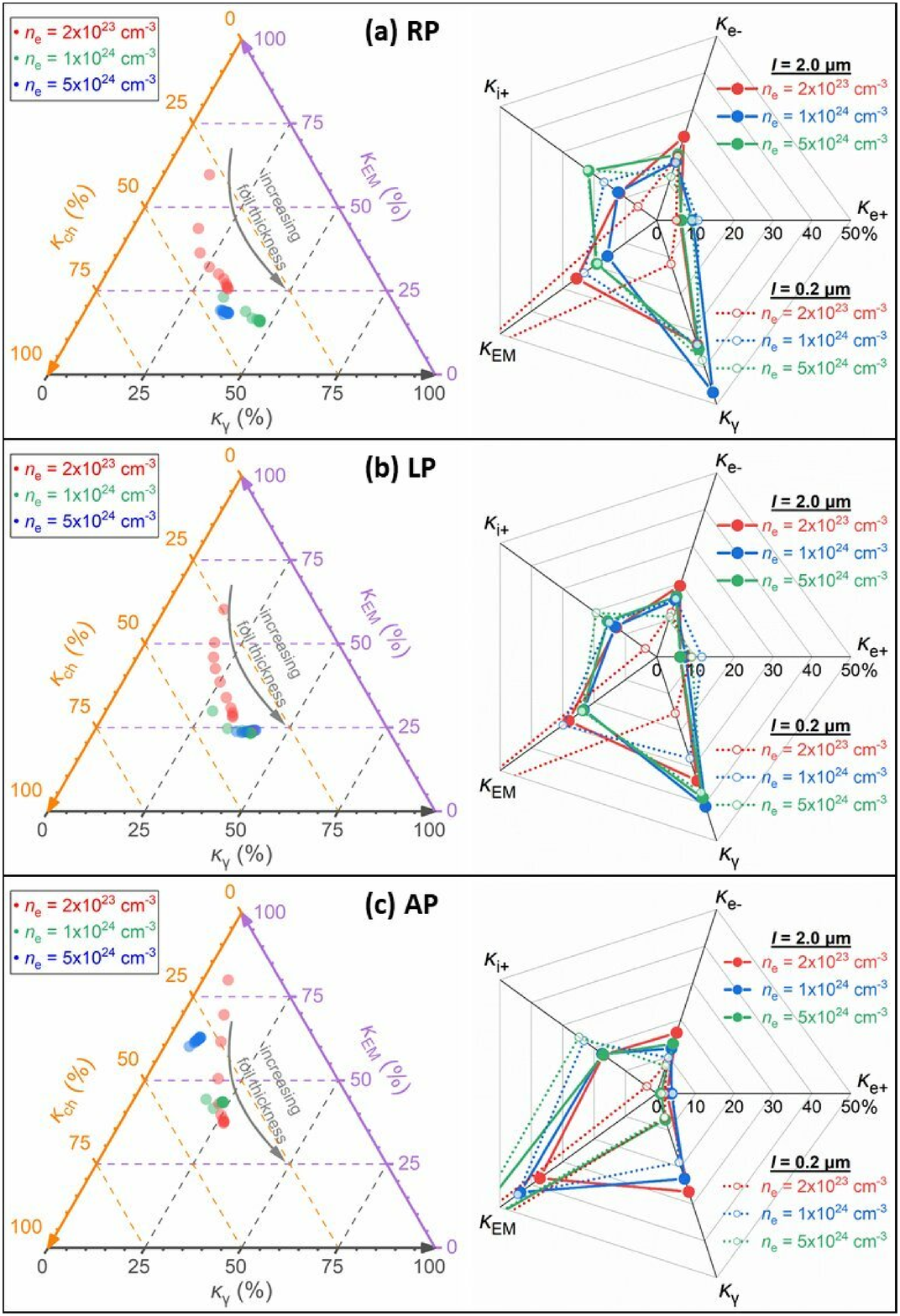}
  \caption{
  (\textit{Left}) Ternary plots of $\kappa_\gamma$, $\kappa_{ch}$ and $\kappa_{EM}$ , for samples with varying electron number density and target thickness. The grey arrow points towards increasing foil thickness. (\textit{Right}) Selected radar charts (solid line for $2 \kern0.2em \mathrm{\upmu m}$ and dotted line for $0.2 \kern0.2em \mathrm{\upmu m}$ thick foil - red for $2 \kern0.1em {\times} \kern0.1em 10^{23} \kern0.2em \mathrm{cm^{-3}}$, blue for $1 \kern0.1em {\times} \kern0.1em 10^{24} \kern0.2em \mathrm{cm^{-3}}$ and green for $5 \kern0.1em {\times} \kern0.1em 10^{24} \kern0.2em \mathrm{cm^{-3}}$ electron number density) of $\kappa_\gamma$, $\kappa_{e+}$, $\kappa_{e-}$, $\kappa_{i+}$ and $\kappa_{EM}$. The cases (\textit{a}), (\textit{b}) and (\textit{c}) correspond to a RP, a LP and an AP laser, respectively.
  }
\label{fig:Ternary}
\end{figure}

 Unavoidably, interaction of a laser field with matter results in transformation of a laser energy fraction to particle energy. The dependency of $\kappa$ on the electron number density and target thickness can be seen in figure \ref{fig:Ternary}, where the direction of the grey arrow on the figure indicates increasing thickness. For both RP and LP lasers, increased laser to all particles energy conversion efficiency, $\kappa_{tot} = \kappa_\gamma + \kappa_{e+} + \kappa_{e-} + \kappa_{i+}$, occurs for thicker and denser targets, ${\sim} \kern0.1em 80 \kern0.2em \%$ and ${\sim} \kern0.1em 85 \kern0.2em \%$ for RP and LP lasers, respectively. For thinner and low density targets the particles obtain only ${\sim} \kern0.1em 40 \kern0.2em \%$ of the laser energy for both RP and LP lasers (within the parameters ranges examined). For an AP laser for thin and low density targets $\kappa_{tot}$ is approximately half compared to RP and LP lasers. For an AP laser (in contrary to the continuously increasing $\kappa_{tot}$ behaviour for RP and LP lasers) beyond of an optimal thickness-density combination $\kappa_{tot}$ starts decreasing for thicker and denser targets (maximum is ${\sim} \kern0.1em 60 \kern0.2em \%$), in connection with the inefficient target cavity formation (see subsection \ref{ElectronEvolution}) and increasing laser back-reflection as electron number density increases.

\par In general, ions being heavier than electrons, they do not produce high energy \textgamma-photons. However, they indirectly affect the \textgamma-photon spectrum. Their contribution arises from the amount of the laser energy transferred to them, consequently reducing electron energy and therefore what can otherwise be converted to \textgamma-photons. For all polarisation cases $\kappa_{i+}$ increases with increasing electron number density up-to an optimum value and then decreases for thicker targets \citep{2004_EsirkepovTZ, 2008_KlimoO, 2008_RobinsonAPL, 2016_BulanovSS}. Therefore, although thin targets can be dense enough to convert a large fraction of laser energy to particle energy, that energy goes primarily to ions. For thick targets, although more laser energy is converted to particle energy by increasing the electron number density, since $\kappa_{i+}$ also increases, it competes with what is converted to $\kappa_\gamma$, $\kappa_{e-}$ and $\kappa_{e+}$, forbidding the optimum of those particles to exist at extremely high electron number density values. For optimal thickness and density combinations, for all three polarisations, $\kappa_{i+}$ reaches ${\sim} \kern0.1em 25 \kern0.2em \%$.

\par Where $\kappa_{i+}$ is not efficient, $\kappa_{e-}$ and $\kappa_{e+}$ cover the imbalance. For all laser polarisation modes, if the electron number density is extremely low then the laser pulse propagates through the target. Alternatively, if the target is thick enough then most of the laser energy is absorbed, resulting in enhanced $\kappa_{e-}$. For RP and LP lasers $\kappa_{e-}$ is ${\sim} \kern0.1em 20 \kern0.2em \%$, while for an AP laser it is ${\sim} \kern0.1em 15 \kern0.2em \%$ at optimum target parameters. Some slow $\kappa_{e-}$ increase for extremely high electron number densities is due to less accurate resolution of the relativistically corrected skin depth, although the increase is insignificantly small to alter the conclusion of the other particle species at that density. For RP and LP lasers, a high $\kappa_{e-}$ also occurs for thin targets in regions where $\kappa_{i+}$ is not efficient, due to electron capture by the laser field \citep{Wang2001}.

\par Although for an ${\sim} \kern0.1em 80 \kern0.2em \mathrm{PW}$ laser a significant number of $\mathrm{e^- \mbox{-} e^+}$ pairs is generated, their number is still relatively low (approximately 50 times lower) compared to the number of electrons contained in the target prior the laser-foil interaction. However, those $\mathrm{e^- \mbox{-} e^+}$ pairs are generated in regions of ultra-intense fields, therefore, stronger heated compared to the electrons in the periphery of the target cavity. The $\mathrm{e^- \mbox{-} e^+}$ pairs are more probable to originate from \textgamma-photons of higher energy. Therefore, $\kappa_{e+}$ is a combination of the energy they obtain from the Breight-Wheeler process, and to the energy due to acceleration/deceleration from the laser field. The thickness-density contour of $\kappa_{e+}$ has partially topological similarities with that of $\kappa_{i+}$, meaning that positrons are affected by the laser field in a similar manner to ions. The $\kappa_{e+}$ for RP and LP lasers reaches ${\sim} \kern0.1em 10 \kern0.2em \%$, while it is approximately half for an AP laser. In contrary to electrons, positrons cannot obtain high $\kappa_{e+}$ for under-dense thick targets because of their low generated number at these parameter values.

\par By combining the laser to all charged particles energy conversion efficiency, $\kappa_{ch} = \kappa_{e+} + \kappa_{e-} + \kappa_{i+}$, we conclude a maximum value of ${\sim} \kern0.1em 45 \kern0.2em \%$ that slowly increases by increasing target thickness, as shown in figure \ref{fig:Ternary}. On the other hand, the figure exhibits a steep increase of $\kappa_\gamma$ for increasing target thickness, where the maximum values are mentioned in subsection \ref{GphotonPositron}. For a LP laser, the topology of $\kappa_\gamma$ thickness-density contour is in agreement with that of $\kappa_{tot}$, being maximised for thick and dense targets. On the other hand, for the RP laser, although the $\kappa_\gamma$ thickness-density contour resembles that of the LP laser for most thickness-density combinations, maximum is observed at an electron number density of $1.2 \kern0.1em \times \kern0.1em 10^{24} \kern0.2em \mathrm{cm^{-3}}$. This local maxima is due to the different rate of energy transfer to ions, where for a RP laser it is lower at that electron number density value. In addition, $\kappa_{tot}$ is slightly higher for the RP laser at thicker and denser targets, further enhancing the local maxima of $\kappa_\gamma$. For an AP laser, the $\kappa_\gamma$ has an optimal electron number density at $5 \kern0.1em \times \kern0.1em 10^{23} \kern0.2em \mathrm{cm^{-3}}$ since the lack of $E_x$ requires a lower electron number density target for efficient laser-target coupling. For more accurate $\kappa$ for each particle species at the extreme thickness-density values one is referred to the right side of figure \ref{fig:Ternary}.


\subsection{Dependency on the Laser Power} \label{PowerScaling}

\par As we have shown for an ${\sim} \kern0.1em 80 \kern0.2em \mathrm{PW}$ RP laser, the $\kappa_\gamma$ is ${\sim} \kern0.1em 47 \kern0.2em \%$ for targets thicker than $2 \kern0.2em \mathrm{\upmu m}$ and an electron number density of $1.2 \times 10^{24} \kern0.2em \mathrm{cm^{-3}}$. A consequent question arises on why the choice of ${\sim} \kern0.1em 80 \kern0.2em \mathrm{PW}$ is made and what is the effect of altering the laser power. To address that topic, the simulations for a RP laser were extended in the power range of $1 \kern0.2em \mathrm{PW} \leqslant P \leqslant 300 \kern0.2em \mathrm{PW}$, where the electron number density was varying in the range $10^{23} \kern0.2em \mathrm{cm^{-3}} \leqslant n_e \leqslant 10^{24} \kern0.2em \mathrm{cm^{-3}}$. As per the results of section \ref{MappingCE}, the $\kappa_\gamma$ varies insignificantly as decreasing the electron number density from $1.2 \times 10^{24} \kern0.2em \mathrm{cm^{-3}}$ to $10^{24} \kern0.2em \mathrm{cm^{-3}}$.

\par
\begin{figure}
  \centering
  \includegraphics[width=1.00\linewidth]{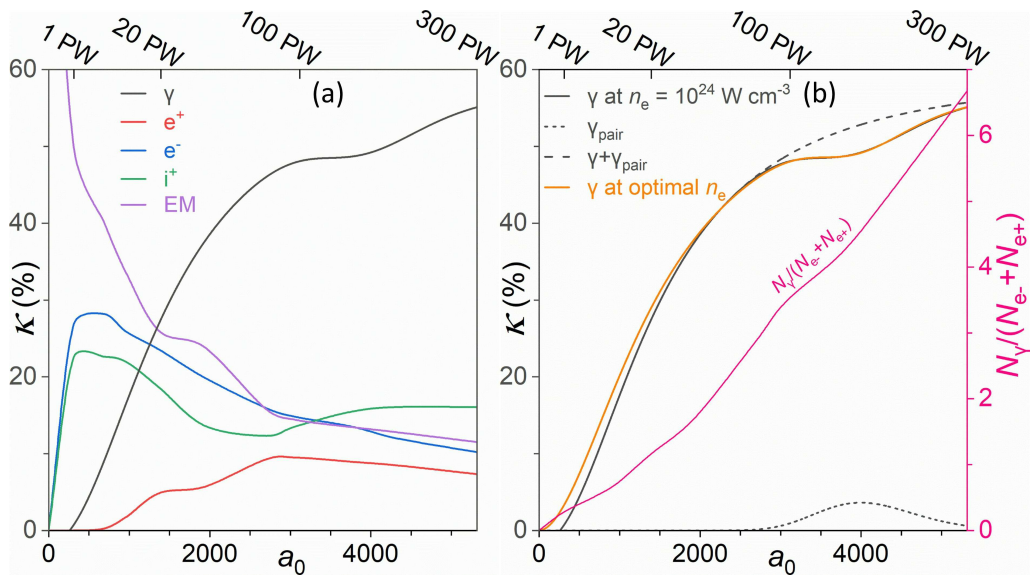}
  \caption{
  (\textit{a}) $\kappa_\gamma$ (black line), $\kappa_{e+}$ (red line), $\kappa_{e-}$ (blue line), $\kappa_{i+}$ (green line) and $\kappa_{EM}$ (purple line) as a function of $a_0$ for a RP $\lambda^3$-laser and an electron number density of $10^{23} - 10^{24} \kern0.2em \mathrm{cm^{-3}}$. (\textit{b - left axis}) $\kappa_\gamma$ fitted with equation \eqref{eq:fit} for an electron number density of $10^{23} - 10^{24} \kern0.2em \mathrm{cm^{-3}}$ (black solid line) and at the optimum electron number density at each power (orange line). The fitted curve is the difference of a ``Logistic'' (long-dashed black line) and a ``LogNormal'' function (short-dashed black line), as defined in the text. (\textit{b - right axis}) The ratio of the \textgamma-photon number over the sum of electron and positron number as a function of $a_0$ for an electron number density of $10^{23} - 10^{24} \kern0.2em \mathrm{cm^{-3}}$.
  }
\label{fig:Power}
\end{figure}

 Let us consider the case where the electron number density is fixed at $10^{24} \kern0.2em \mathrm{cm^{-3}}$ and the laser power varies. The $\kappa$ of each species is shown in figure \ref{fig:Power}, where $a_0 \approx 307$ for $1 \kern0.2em \mathrm{PW}$, while $a_0 \approx 5318$ for $300 \kern0.2em \mathrm{PW}$. The $\kappa_\gamma$, $\kappa_{e+}$, $\kappa_{e-}$ and $\kappa_{i+}$ are shown with the black, red, blue and green continuous lines, respectively, while the percentage of the laser energy remaining as electromagnetic energy, $\kappa_{EM}$, is shown by the purple continuous line.

\par From the purple line in figure \ref{fig:Power} one can observe that at low laser power the laser cannot be efficiently absorbed by the target and it is mostly reflected, since at low power the skin depth does not have significant relativistic increase. However, by increasing the laser power to $20 \kern0.2em \mathrm{PW}$, corresponding to $a_0 \sim 1400$, ${\sim} \kern0.1em 75 \kern0.2em \%$ of the laser energy is absorbed by the target. By further increasing the power up-to $300 \kern0.2em \mathrm{PW}$ the percentage of the laser energy absorbed increases, although with a lower rate as power increases and eventually saturating at ${\sim} \kern0.1em 10 \kern0.2em \%$.

\par At ${\sim} \kern0.1em 20 \kern0.2em \mathrm{PW}$ the $\kappa_\gamma$,  $\kappa_{e-}$ and  $\kappa_{i+}$ becomes equally important. At $P \lessapprox 5 \kern0.2em \mathrm{PW}$, most of the laser energy is transferred to electrons and ions, with \textgamma-photons and positrons obtaining an insignificantly low laser energy fraction. However, the picture reverses for $P \gtrapprox 20 \kern0.2em \mathrm{PW}$, where $\kappa_{i+}$ saturates at ${\sim} \kern0.1em 15 \kern0.2em \%$. The $\kappa_{e-}$ also exhibits a plateau region at $1 \kern0.2em \mathrm{PW} \lessapprox P \lessapprox 5 \kern0.2em \mathrm{PW}$, after which, $\kappa_{e-}$ continuously decreases for increasing laser power and eventually saturating at ${\sim} \kern0.1em 10 \kern0.2em \%$. The $\kappa_{e+}$ continuously increases for laser power up to ${\sim} \kern0.1em 80 \kern0.2em \mathrm{PW}$, where after obtaining a maximum value of ${\sim} \kern0.1em 9 \kern0.2em \%$ it decreases to ${\sim} \kern0.1em 5 \kern0.2em \%$ for higher power values.

\par The trend of $\kappa_\gamma$ in figure \ref{fig:Power} changes at ${\sim} \kern0.1em 80 \kern0.2em \mathrm{PW}$ power. Since $\kappa_{e+}$, $\kappa_{e-}$ and $\kappa_{i+}$ all saturate for increasing power, then $\kappa_\gamma$ unavoidably also saturates, where the sum of $\kappa_{e+}$, $\kappa_{e-}$ and $\kappa_{i+}$ suggests a $\kappa_\gamma$ saturation at ${\sim} \kern0.1em 60 \kern0.2em \%$. Therefore, we treat the $\kappa_\gamma$ function as the difference of a ``Logistic'' and a ``LogNormal'' function, given respectively by the left and right parts of equation
\begin{equation}
 \kappa_\gamma = A_2 + \frac{A_1-A_2}{1+\left( x/x_0 \right)^p} - \frac{A_3}{w x} \exp \left\{ -\frac{\left[ \ln(x / x_c) \right]^2}{2 w^2} \right\} .
\label{eq:fit}
\end{equation}
Fitting of equation \eqref{eq:fit} to $\kappa_\gamma$ as shown in figure \ref{fig:Power}(b) gives $A_1 \approx -1.75$, $A_2 \approx 59.8$, $p \approx 2.05$, $x_0 \approx 1463$, $A_3 \approx 2213$, $w \approx -0.151$ and $x_c \approx 4088$.

\par The ``Logistic'' function [black dashed line in figure \ref{fig:Power}(b)] explains the expected $\kappa_\gamma$ saturation for an increasing laser power. The parameter $A_2$ suggests $\kappa_\gamma$ saturation at ${\sim} \kern0.1em 59.8 \kern0.2em \%$, while the parameter $A_1$ suggests that at an electron number density of $10^{24} \kern0.2em \mathrm{cm^{-3}}$ no \textgamma-photons can be produced for a laser power of ${\sim} \kern0.1em 0.7 \kern0.2em \mathrm{PW}$. The ``LogNormal'' function [black dotted line in figure \ref{fig:Power}(b)], having a negative sign, suggests that a \textgamma-photon population is lost to $\mathrm{e^- \mbox{-} e^+}$ pairs, where their contribution becomes most significant for an ${\sim} \kern0.1em 177 \kern0.2em \mathrm{PW}$ as suggested by the parameter $x_c$.

\par By repeating the analysis described above for electron number densities in the range $10^{23} \kern0.2em \mathrm{cm^{-3}} \leqslant n_e \leqslant 10^{24} \kern0.2em \mathrm{cm^{-3}}$ we find the optimal electron density value at each power for maximising $\kappa_\gamma$, plotted by the orange line in figure \ref{fig:Power}(b). The trend suggests that a $1 \kern0.2em \mathrm{PW}$ is sufficient for a $\kappa_\gamma$ of ${\sim} \kern0.1em 3 \kern0.2em \%$. The density-power contour suggests that $\kappa_\gamma$ is strongly dependent on the electron number density at low laser power, optimal at $2 \times 10^{23} \kern0.2em \mathrm{cm^{-3}}$ for a $1 \kern0.2em \mathrm{PW}$ laser. By increasing the laser power, denser targets are required to give the peak $\kappa_\gamma$, although the density dependency becomes less prominent as power increases.

\par The pink line on the right side of figure \ref{fig:Power}(b) shows the ratio of \textgamma-photon number produced to the sum of electron and positron number as a function of $a_0$. The line exhibits an approximately linearly increasing trend, suggesting that at higher laser powers each electron/positron can emit \textgamma-photons several times by the end of the simulation. For an ${\sim} \kern0.1em 80 \kern0.2em \mathrm{PW}$ laser, each electron/positron emits \textgamma-photons approximately three times.


\section{\textgamma-flash Interaction with High-Z target} \label{FLUKA}


\par In order to examine the effect of the \textgamma-flash described in section \ref{Gflash} on a secondary, high-Z target, we perform MC simulations using the FLUKA code \citep{2015_BattistoniG, 2014_BoehlenTT} and its graphical interface FLAIR\citep{2009_VlachoudisV}. In addition to \textgamma-photons, the effects of the charged PIC-produced particles with the secondary target are also investigated. The PIC output particles (type, position, momentum and weight) are imported to FLUKA as primary particles. The secondary target is modelled as a $10\kern0.2em \mathrm{mm}$ thick disk of $100\kern0.2em \mathrm{mm}$ diameter and it is located at $0.1\kern0.2em \mathrm{mm}$ from the focal spot coordinates. The large acceptance angle covered by the secondary target allows to intercept almost all PIC generated particles in the forward direction. Natural lead, Pb, is chosen as material for the disk because of its high cross section for pair production and photonuclear interactions for energies considered. For the simulations, the FLUKA PRECISIO defaults are used. Additionally, the electromagnetic transport thresholds are set at $0.1 \kern0.2em \mathrm{MeV}$, the photonuclear and electronuclear interactions are enabled, as well as the evaporation of heavy fragments and nuclear coalescence.


\begin{figure}
  \centering
  \includegraphics [width=0.75\linewidth]{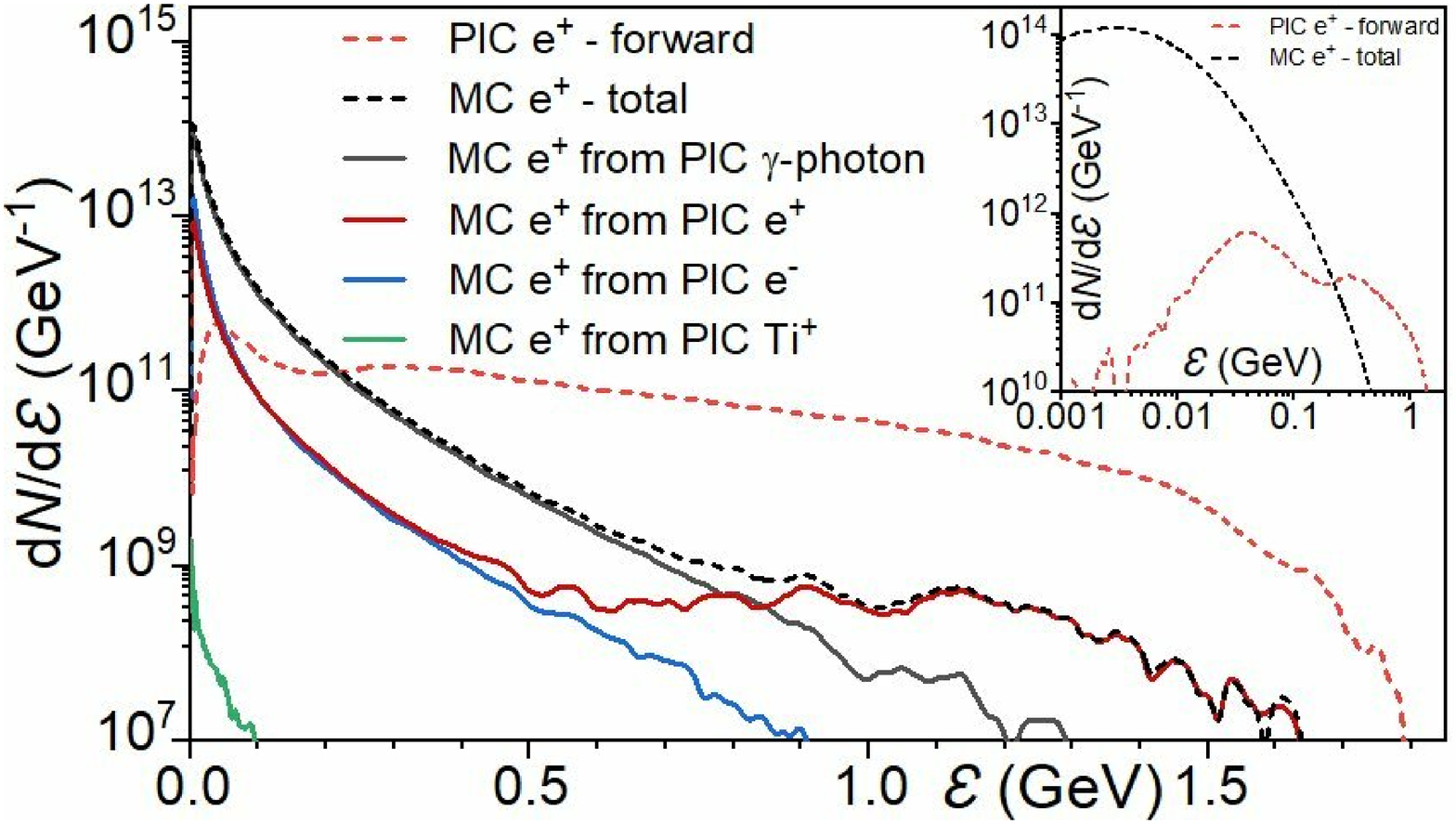}
  \caption{The figure shows, per $\lambda^3$-pulse, the energy spectrum of the PIC positrons moving in the forward direction (red dashed), along with the positron spectra from the MC simulations in total (black dashed line), and separated per producing species [\textgamma-photons (black), electrons (blue), positrons (red) and titanium ions (green)].} \label{fig:FLUKApositrons}
\end{figure}

\par Figure \ref{fig:FLUKApositrons} shows that the PIC generated positrons moving in the forward direction, exhibiting a rather flat spectrum with a temperature of ${\sim} \kern0.1em 0.4 \kern0.2em \mathrm{GeV}$. The figure overplots the spectra of positrons escaping the secondary target in the forward direction, obtained from the MC simulations, integrated (black dashed line) and separated per each primary particle species (solid lines), namely \textgamma-photons, electrons, positrons and titanium ions. From figure \ref{fig:FLUKApositrons} it is seen that the largest number of positrons (${\sim} \kern0.1em 81.4 \kern0.2em \%$) is produced by \textgamma-photons and that the most energetic positrons are those directly created in the PIC simulations. 

\par Positrons produced by PIC \textgamma-photons and electrons have a temperature of ${\sim} \kern0.1em 0.1 \kern0.2em \mathrm{GeV}$. The positron population exhibits two temperatures, the first of ${\sim} \kern0.1em 0.1 \kern0.2em \mathrm{GeV}$ corresponding to those generated in the lead target, and the second at higher temperature corresponding to PIC generated positrons. The low temperature positrons are generated via $\mathrm{e^- \mbox{-} e^+}$ pair production from Bremsstrahlung \textgamma-photons. The positron spectra after the secondary target is shifted towards lower energy with respect to the PIC-produced positrons, while their total number is increased by approximately an order of magnitude.

\begin{figure}
  \centering
  \includegraphics [width=0.9\linewidth]{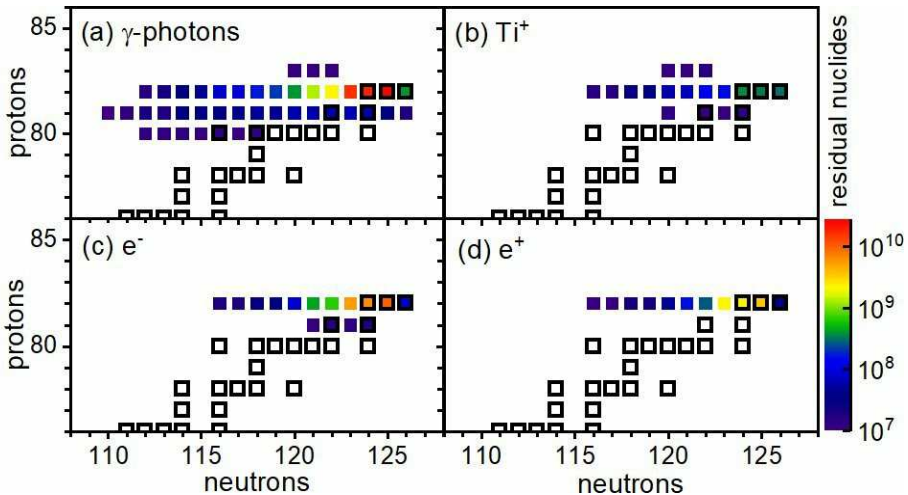}
  \caption{Chart of residual nuclides obtained from MC simulation per $\lambda^3$-pulse and separated per each PIC particle species. Stable nuclides are highlighted with a box.}
  \label{fig:FLUKAactivation}
\end{figure}

\par MC simulations also allow to estimate the number of stable and unstable nuclides generated in the lead target. Figure \ref{fig:FLUKAactivation} shows the chart of the produced nuclides focused around the lead position and separated per each PIC particle species. Stable nuclides are highlighted with a box. Most of the residual nuclides are produced through photonuclear interactions, either directly by primary (PIC) \textgamma-photons or indirectly by secondary (Bremsstrahlung from fast electrons/positrons) \textgamma-photons. In our \textgamma-photon energy region of interest, the Giant Dipole Resonance (GDR) photonuclear process dominates since it has the highest integrated cross section, peaking at ${\sim} \kern0.1em 13.6 \kern0.2em \mathrm{MeV}$ \textgamma-photons. Apart of photonuclear interactions, residual nuclides can be also produced by nucleus-nucleus interactions and/or electronuclear interactions.

\par  One of the most abundant generated lead isotopes is \ce{^{203}_{82}Pb} where ${\sim} \kern0.1em 10^9$ nuclides are produced with a half life of ${\sim} \kern0.1em 52 \kern0.2em \mathrm{h}$. Its direct decay to \ce{^{203}_{81}Tl} (stable) through electron capture and it does not emit any hadrons. In addition, photons of ${\sim} \kern0.1em 279.2 \kern0.2em \mathrm{keV}$ are emitted which are particularly suitable for medical imaging \citep{2014_AzzamA}. The second most abundant isotope produced is thalium, with \ce{^{201}_{81}Tl} (${\sim} \kern0.1em 10^8$ nuclides) being historically used extensively for nuclear medicine \citep{1999_TadamuraE} due to its decay to \ce{^{201}_{80}Hg} (stable) through electron capture with a half-life of ${\sim} \kern0.1em 73 \kern0.2em \mathrm{h}$. 


\section{Summary and Conclusions}

\par In this work we study the highly efficient \textgamma-photon generation through the ultra-intense laser and solid target interaction.  We employ the $\lambda^3$ regime, where a single-cycle laser ${\sim} \kern0.1em 80 \kern0.2em \mathrm{PW}$ laser pulse is focused to a ${\sim} \kern0.1em \lambda/2$ diameter sphere. The benefit of the $\lambda^3$ regime is that it provides the highest intensity achievable at a given laser power, in expense of the least energy. In this paper we study the interaction of a $\lambda^3$-laser with matter in the QED regime, where a copious number of \textgamma-photons and $\mathrm{e^- \mbox{-} e^+}$ pairs is generated. The QED processes are studied by use of the 3D EPOCH PIC code. The $\lambda^3$-laser fields are imported into EPOCH after calculated independently through our code developed.

\par Our work examines the laser-target interaction under RP, LP and AP lasers. A multi-parametric study is presented, where the variables include the target thickness and electron number density. It is found that the optimal $\kappa_\gamma$ reaches ${\sim} \kern0.1em 47 \kern0.2em \%$ and it occurs for a RP laser at a target thickness of $2 \kern0.2em \mathrm{\upmu m}$ and an electron number density of $1.2 \kern0.1em \times \kern0.1em 10^{24} \kern0.2em \mathrm{cm^{-3}}$. For the same target parameters, the LP and AP lasers results in a $\kappa_\gamma$ of ${\sim} \kern0.1em 40 \kern0.2em \%$ and ${\sim} \kern0.1em 20 \kern0.2em \%$, respectively. At the optimal target variables, the LP lasers gives a $\kappa_\gamma$ of ${\sim} \kern0.1em 42 \kern0.2em \%$, while the AP laser gives a $\kappa_\gamma$ of ${\sim} \kern0.1em 29 \kern0.2em \%$.

\par The significantly higher $\kappa_\gamma$ for the RP laser is due to the dominance of the longitudinal, $E_x$ field, that increases the coupling of the laser to the target. For the LP laser the $E_x$ is smaller, where for the AP laser is absent. The $E_x$ assists in the formation of a target cavity, where the cavity propagation performs a different propagation depending on the laser polarisation mode. Interference of the reflected/diffracted laser field inside the cavity results in an instantaneous intensity as high as ${\sim} \kern0.1em 8.8  \kern0.1em \times \kern0.1em 10^{25} \kern0.2em \mathrm{W cm^{-2}}$, approximately one order of magnitude higher than the intensity expected on focus.

\par The directionality of electrons at several instances is identified, resulting in several high energy electron groups directed at ${\sim} \kern0.1em 0 \degree$, ${\sim} \kern0.1em 180 \degree$ and ${\sim} \kern0.1em 60 \degree$ for a RP laser. Those electrons are connected to the \textgamma-photon directionality, being at the same angles. The ultra-high intensities employed, result in not only a prolific \textgamma-photon generation but unavoidably to also $\mathrm{e^- \mbox{-} e^+}$ pair generation through the multi-photon Breit-Wheeler process. The generation positions of $\mathrm{e^- \mbox{-} e^+}$ pairs is identified to overlap the regions of high-energy \textgamma-photons.

\par At a time of ${\sim} \kern0.1em \lambda/(2 c)$ after the peak of the laser pulse reaches the focal spot, the \textgamma-photons expand radially in a ballistic fashion without significant losses to $\mathrm{e^- \mbox{-} e^+}$ pairs. The \textgamma-photons expand within a spherical shell where the FWHM of their energy density is approximately equal to the laser wavelength, similar to the laser-foil interaction time. The expanding spherical shell for the RP, LP and AP laser results in a \textgamma-flash of ${\sim} \kern0.1em 31 \kern0.2em \mathrm{PW}$, ${\sim} \kern0.1em 28 \kern0.2em \mathrm{PW}$ and ${\sim} \kern0.1em 13 \kern0.2em \mathrm{PW}$, respectively. Although a preferred directionality exists for the \textgamma-photons, the radiant intensity of the population at ${\sim} \kern0.1em 60 \degree$ is less significant due to its large solid angle, in contrary to \textgamma-photons at ${\sim} \kern0.1em 0 \degree$ and ${\sim} \kern0.1em 180 \degree$.

\par Our analysis is also extended on varying the laser power in the range $1 \kern0.2em \mathrm{PW} \leqslant P \leqslant 300 \kern0.2em \mathrm{PW}$. We demonstrate that $\kappa_\gamma$ sharply increases up to ${\sim} \kern0.1em 80 \kern0.2em \mathrm{PW}$, while \textgamma-photons become the dominant species above ${\sim} \kern0.1em 20 \kern0.2em \mathrm{PW}$. For low laser powers strong dependency of the $\kappa_\gamma$ exists on the electron number density, where the optimal electron number density increases approximately linearly with $a_0$. For higher power values this dependency becomes less important. When increasing the laser power then the $\kappa_\gamma$ increases, saturating at ${\sim} \kern0.1em 60 \kern0.2em \%$. A $\kappa_\gamma$ discontinuity exists centred at ${\sim} \kern0.1em 177 \kern0.2em \mathrm{PW}$, attributed to the \textgamma-photon conversion to $\mathrm{e^- \mbox{-} e^+}$ pairs, while as power further increases the \textgamma-photon reduction is compensated by further \textgamma-photon emission by positrons, in the same manner as by electrons. In addition, as laser power increases then the number of \textgamma-photon emission from each electron/positron also increases in an approximately linear fashion with $a_0$.

\par Finally, for the RP laser, the interaction of the PIC generated particles interacting with a high-Z target is studied by MC simulations. The spectra of each particle species escaping the secondary target are obtained. The PIC spectra are substantially altered by the interaction with the high-Z target. The \textgamma-flash interaction with the secondary target also results in significant production of radioactive nuclides, whose yields are estimated. Hence, the coupling of PIC and MC simulations provides a powerful tool for further investigating the laser interaction with matter.


\hfill \break
\par The authors would like to acknowledge useful communication with Dr. D. Khikhlukha and K. Lezhnin. This work is supported by the projects High Field Initiative (CZ.02.1.01/0.0/0.0/15\_003/0000449) from the European Regional Development Fund and ``e-INFRA CZ'' (ID:90140) from the Ministry of Education, Youth and Sports of the Czech Republic. CPR would like to acknowledge funding from EPSRC, grant no.  EP/V049461/1. The EPOCH code is in part funded by the UK EPSRC grants EP/G054950/1, EP/G056803/1, EP/G055165/1 and EP/M022463/1.


\appendix

\section{}\label{appA}

The imaginary part of the integrands used for the electric and magnetic field calculation \citep{2015_JeongTM} of a LP laser, as used in our Fortran code. The definition of symbols is found in section \ref{Lambda3}. The electric field integrands are

\begin{equation}
I_{Ex-L} = -A^2 B \cos(\phi) X \cos(F) + [A^2 cos(\phi) - A^2 \cos(\phi) X] \sin(F)
\label{LEx}
\end{equation}

\begin{equation}
I_{Ey-L} = -A^2 B \cos(\phi) Y \cos(F) + [A - A^2 \cos(\phi) Y] \sin(F)
\label{LEy}
\end{equation}

\begin{equation}
I_{Ez-L} = -A^2 B \cos(\phi) Z \cos(F) - [A^2 \cos(\phi) Z] \sin(F)
\label{LEz}
\end{equation}

\par The magnetic field integrands are

\begin{equation}
I_{Bx-L} = A B Z \cos(F) +A Z \sin(F)
\label{LBx}
\end{equation}

\begin{equation}
I_{By-L} = -A^2 B \cos(\phi) Z \cos(F) - [A^2 \cos(φ) Z] \sin(F)
\label{LBy}
\end{equation}

\begin{equation}
I_{Bz-L} = [A^2 B \cos(\phi) Y - A B X] \cos(F) +[A^2 \cos(\phi) Y - A X] \sin(F)
\label{LBz}
\end{equation}

\section{}\label{appB}

The imaginary part of the integrands used for the electric and magnetic field calculation \citep{2018_JeongTM} of a RP laser, as used in our Fortran code. The definition of symbols is found in section \ref{Lambda3}. The electric field integrands are

\begin{equation}
I_{Ex-R} = A^2 B X \sin(F) -[A^2 X - A^2] \cos(F)
\label{REx}
\end{equation}

\begin{equation}
I_{Ey-R} = A^2 B Y \sin(F) -[A^2 Y - A \cos(\phi)] \cos(F)
\label{REy}
\end{equation}

\begin{equation}
I_{Ez-R} = A^2 B Z \sin(F) -[A^2 Z - A \sin(\phi)] \cos(F)
\label{REz}
\end{equation}

\par The magnetic field integrands are

\begin{equation}
I_{Bx-R} = 0
\label{RBx}
\end{equation}

\begin{equation}
I_{By-R} = A [ \sin(\phi) X - A Z ] \cos(F) - A B [ \sin(\phi) X - A Z ] \sin(F)
\label{RBy}
\end{equation}

\begin{equation}
I_{Bz-R} = -A [ \cos(\phi) X - A Y ] \cos(F) + A B [ \cos(\phi) X - A Y ] \sin(F)
\label{RBz}
\end{equation}


\bibliographystyle{jpp}
\bibliography{jpp2021_bib}

\providecommand{\noopsort}[1]{}\providecommand{\singleletter}[1]{#1}%
\begin{thebibliography}{88}
\expandafter\ifx\csname natexlab\endcsname\relax\def\natexlab#1{#1}\fi
\def\au#1{#1} \def\ed#1{#1} \def\yr#1{#1}\def\at#1{#1}\def\jt#1{\textit{#1}}
  \def\bt#1{#1}\def\bvol#1{\textbf{#1}} \def\vol#1{#1} \def\pg#1{#1}
  \def\publ#1{#1}\def\arxiv#1{#1}\def\org#1{#1}\def\st#1{\textit{#1}}

\bibitem[199(1998)]{1998_iaea}
 \yr{1998} {\em {Accelerator Driven Systems:Energy Generation and Transmutation
  of Nuclear Waste: Status Report}\/}. {\em TECDOC Series\/} 985.
  \publ{Vienna: INTERNATIONAL ATOMIC ENERGY AGENCY}.

\bibitem[Aharonian {\em et~al.\/}(2021)Aharonian, An, Axikegu, Bai, Bai, Bao,
  Bastieri, Bi, Bi \& Cai]{2021_AharonianF}
{\sc \au{Aharonian, F.}, \au{An, Q.}, \au{Axikegu}, \au{Bai, L.~X.}, \au{Bai,
  Y.~X.}, \au{Bao, Y.~W.}, \au{Bastieri, D.}, \au{Bi, X.~J.}, \au{Bi, Y.~J.} \&
  \au{Cai, H. et~al.}} \yr{2021}  \at{{Extended Very-High-Energy Gamma-Ray
  Emission Surrounding $\mathrm{PSR \kern0.2em J}0622+3749$ Observed by
  $\mathrm{LHAASO-KM2A}$}}.  \jt{Phys. Rev. Lett.}  \bvol{126},  \pg{241103}.

\bibitem[Aichelin(1991)]{1991_AichelinJ}
{\sc \au{Aichelin, J.}} \yr{1991}  \at{{“Quantum” molecular dynamics—a
  dynamical microscopic n-body approach to investigate fragment formation and
  the nuclear equation of state in heavy ion collisions}}.  \jt{Phys. Reports}
  \bvol{202}~(5),  \pg{233--360}.

\bibitem[April \& Pich\'{e}(2010)]{2010_April}
{\sc \au{April, A.} \& \au{Pich\'{e}, M.}} \yr{2010}  \at{{4$\pi$ Focusing of
  TM01 beams under nonparaxial conditions}}.  \jt{Opt. Express}
  \bvol{18}~(21),  \pg{22128--22140}.

\bibitem[Arber {\em et~al.\/}(2015)Arber, Bennett, Brady, Lawrence-Douglas,
  Ramsay, Sircombe, Gillies, Evans, Schmitz \& Bell]{2015_ArberTD}
{\sc \au{Arber, T.~D.}, \au{Bennett, K.}, \au{Brady, C.~S.},
  \au{Lawrence-Douglas, A.}, \au{Ramsay, M.~G.}, \au{Sircombe, N.~J.},
  \au{Gillies, P.}, \au{Evans, R.~G.}, \au{Schmitz, H.} \& \au{Bell, A. R.
  et~al.}} \yr{2015}  \at{{Contemporary particle-in-cell approach to
  laser-plasma modelling}}.  \jt{Plasma Phys. Control. Fusion}  \bvol{57}~(11),
   \pg{113001}.

\bibitem[Audet {\em et~al.\/}(2021)Audet, Alejo, Calvin, Cunningham, Frazer,
  Nersisyan, Phipps, Warwick, Sarri \& Hafz]{2021_AudetTL}
{\sc \au{Audet, T.~L.}, \au{Alejo, A.}, \au{Calvin, L.}, \au{Cunningham,
  M.~H.}, \au{Frazer, G.R.}, \au{Nersisyan, G.}, \au{Phipps, M.~l},
  \au{Warwick, J~.R.}, \au{Sarri, G.} \& \au{Hafz, N. A. M. et~al.}} \yr{2021}
  \at{{Ultrashort, MeV-scale laser-plasma positron source for positron
  annihilation lifetime spectroscopy}}.  \jt{Phys. Rev. Accel. Beams}
  \bvol{24},  \pg{073402}.

\bibitem[Azzam {\em et~al.\/}(2014)Azzam, Said \& Al-abyad]{2014_AzzamA}
{\sc \au{Azzam, A.}, \au{Said, S.~A.} \& \au{Al-abyad, M.}} \yr{2014}
  \at{{Evaluation of different production routes for the radio medical isotope
  203Pb using TALYS 1.4 and EMPIRE 3.1 code calculations}}.  \jt{Appl. Radiat.
  Isot.}  \bvol{91},  \pg{109--113}.

\bibitem[Bahk {\em et~al.\/}(2004)Bahk, Rousseau, Planchon, Chvykov,
  Kalintchenko, Maksimchuk, Mourou \& Yanovsky]{2004_Bahk}
{\sc \au{Bahk, S.~W.}, \au{Rousseau, P.}, \au{Planchon, T.~A.}, \au{Chvykov,
  V.}, \au{Kalintchenko, G.}, \au{Maksimchuk, A.}, \au{Mourou, G.~A.} \&
  \au{Yanovsky, V.}} \yr{2004}  \at{{Generation and characterization of the
  highest laser intensities (1022 W/cm2)}}.  \jt{Opt. Lett.}  \bvol{29}~(24),
  \pg{2837--2839}.

\bibitem[Battistoni {\em et~al.\/}(2015)Battistoni, Boehlen, Cerutti, Chin,
  Esposito, Fassò, Ferrari, Lechner, Empl \& Mairani]{2015_BattistoniG}
{\sc \au{Battistoni, G.}, \au{Boehlen, T.}, \au{Cerutti, F.}, \au{Chin, P.W.},
  \au{Esposito, L.S.}, \au{Fassò, A.}, \au{Ferrari, A.}, \au{Lechner, A.},
  \au{Empl, A.} \& \au{Mairani, A. et~al.}} \yr{2015}  \at{{Overview of the
  FLUKA code}}.  \jt{Ann. Nucl. Energy}  \bvol{82},  \pg{10--18}.

\bibitem[Bell \& Kirk(2008)]{2008_BellAR}
{\sc \au{Bell, A.~R.} \& \au{Kirk, John~G.}} \yr{2008}  \at{{Possibility of
  Prolific Pair Production with High-Power Lasers}}.  \jt{Phys. Rev. Lett.}
  \bvol{101},  \pg{200403}.

\bibitem[Berestetskii {\em et~al.\/}(1982)Berestetskii, Lifshitz \&
  Pitaevskii]{1982_BerestetskiiVB}
{\sc \au{Berestetskii, V.~B.}, \au{Lifshitz, E.~M.} \& \au{Pitaevskii, L.~P.}}
  \yr{1982} {\em {Quantum Electrodynamics (Second Edition)}\/}, second edition
  edn.  \publ{Oxford: Butterworth-Heinemann}.

\bibitem[Bethe \& Heitler(1934)]{1934_BetheH}
{\sc \au{Bethe, H.} \& \au{Heitler, W.}} \yr{1934}  \at{{On the Stopping of
  Fast Particles and on the Creation of Positive Electrons}}.  \jt{Proc. R.
  Soc. Lond. A}  \bvol{146}~(856),  \pg{83--112}.

\bibitem[Böhle {\em et~al.\/}(2014)Böhle, Kretschmar, Jullien, Kovacs,
  Miranda, Romero, Crespo, Morgner, Simon \& Lopez-Martens]{2014_BohleF}
{\sc \au{Böhle, F.}, \au{Kretschmar, M.}, \au{Jullien, A.}, \au{Kovacs, M.},
  \au{Miranda, M.}, \au{Romero, R.}, \au{Crespo, H.}, \au{Morgner, U.},
  \au{Simon, P.} \& \au{Lopez-Martens, R. et~al.}} \yr{2014}  \at{{Compression
  of {CEP}-stable multi-{mJ} laser pulses down to 4{\hspace{0.167em}}fs in long
  hollow fibers}}.  \jt{Laser Phys. Lett.}  \bvol{11}~(9),  \pg{095401}.

\bibitem[Böhlen {\em et~al.\/}(2014)Böhlen, Cerutti, Chin, Fassò, Ferrari,
  Ortega, Mairani, Sala, Smirnov \& Vlachoudis]{2014_BoehlenTT}
{\sc \au{Böhlen, T.~T.}, \au{Cerutti, F.}, \au{Chin, M. P.~W.}, \au{Fassò,
  A.}, \au{Ferrari, A.}, \au{Ortega, P.~G.}, \au{Mairani, A.}, \au{Sala,
  P.~R.}, \au{Smirnov, G.} \& \au{Vlachoudis, V.}} \yr{2014}  \at{{The FLUKA
  Code: Developments and Challenges for High Energy and Medical Applications}}.
   \jt{Nucl. Data Sheets}  \bvol{120},  \pg{211--214}.

\bibitem[Boris(1970)]{Boris1970}
{\sc \au{Boris, J.~P.}} \yr{1970}  \at{{Relativistic plasma
  simulation-optimization of a hybrid code}}.  \jt{Proceeding of Fourth
  Conference on Numerical Simulations of Plasmas}  \pg{pp. 3--67}.

\bibitem[Budnev {\em et~al.\/}(1975)Budnev, Ginzburg, Meledin \&
  Serbo]{1975_BudnevVM}
{\sc \au{Budnev, V.~M.}, \au{Ginzburg, I.F.}, \au{Meledin, G.V.} \& \au{Serbo,
  V.G.}} \yr{1975}  \at{{The two-photon particle production mechanism. Physical
  problems. Applications. Equivalent photon approximation}}.  \jt{Phys.
  Reports}  \bvol{15}~(4),  \pg{181--282}.

\bibitem[Bulanov {\em et~al.\/}(2016)Bulanov, Esarey, Schroeder, Bulanov,
  Esirkepov, Kando, Pegoraro \& Leemans]{2016_BulanovSS}
{\sc \au{Bulanov, S.~S.}, \au{Esarey, E.}, \au{Schroeder, C.~B.}, \au{Bulanov,
  S.~V.}, \au{Esirkepov, T.~Z.}, \au{Kando, M.}, \au{Pegoraro, F.} \&
  \au{Leemans, W.~P.}} \yr{2016}  \at{{Radiation pressure acceleration: The
  factors limiting maximum attainable ion energy}}.  \jt{Phys. Plasmas}
  \bvol{23}~(5),  \pg{056703}.

\bibitem[Bulanov {\em et~al.\/}(2006)Bulanov, Esirkepov, Kamenets \&
  Pegoraro]{2006_BulanovSS}
{\sc \au{Bulanov, S.~S.}, \au{Esirkepov, T.~Z.}, \au{Kamenets, F.~F.} \&
  \au{Pegoraro, F.}} \yr{2006}  \at{{Single-cycle high-intensity
  electromagnetic pulse generation in the interaction of a plasma wakefield
  with regular nonlinear structures}}.  \jt{Phys. Rev. E}  \bvol{73},
  \pg{036408}.

\bibitem[Bulanov {\em et~al.\/}(2015)Bulanov, Esirkepov, Kando, Koga, Kondo \&
  Korn]{2015_BulanovSV}
{\sc \au{Bulanov, S.~V.}, \au{Esirkepov, T.~Z.}, \au{Kando, M.}, \au{Koga, J.},
  \au{Kondo, K.} \& \au{Korn, G.}} \yr{2015}  \at{{On the problems of
  relativistic laboratory astrophysics and fundamental physics with super
  powerful lasers}}.  \jt{Plasma Phys. Rep.}  \bvol{41},  \pg{1--51}.

\bibitem[Chen {\em et~al.\/}(2011)Chen, Meyerhofer, Wilks, Cauble, Dollar,
  Falk, Gregori, Hazim, Moses \& Murphy]{2011_ChenH}
{\sc \au{Chen, H.}, \au{Meyerhofer, D.~D.}, \au{Wilks, S.~C.}, \au{Cauble, R.},
  \au{Dollar, F.}, \au{Falk, K.}, \au{Gregori, G.}, \au{Hazim, A.}, \au{Moses,
  E.~I.} \& \au{Murphy, C. D. et~al.}} \yr{2011}  \at{{Towards laboratory
  produced relativistic electron–positron pair plasmas}}.  \jt{High Energy
  Density Phys.}  \bvol{7}~(4),  \pg{225--229}.

\bibitem[Cheng {\em et~al.\/}(2015)Cheng, Zhou, Xia, Li, Yang \&
  Zhou]{2015_Cheng}
{\sc \au{Cheng, Z.}, \au{Zhou, Y.}, \au{Xia, M.}, \au{Li, W.}, \au{Yang, K.} \&
  \au{Zhou, Y.}} \yr{2015}  \at{{Tight focusing of the azimuthally polarized
  light beam for a sharper spot}}.  \jt{Opt. Laser Technol.}  \bvol{73},
  \pg{77--81}.

\bibitem[Compton(1923)]{1923_ComptonAH}
{\sc \au{Compton, A.~H.}} \yr{1923}  \at{{A Quantum Theory of the Scattering of
  X-rays by Light Elements}}.  \jt{Phys. Rev.}  \bvol{21},  \pg{483--502}.

\bibitem[Danson {\em et~al.\/}(2019)Danson, Haefner, Bromage, Butcher,
  Chanteloup, Chowdhury, Galvanauskas, Gizzi, Hein \& Hillier]{2019_DansonC}
{\sc \au{Danson, C.~N.}, \au{Haefner, C.}, \au{Bromage, J.}, \au{Butcher, T.},
  \au{Chanteloup, J.~F.}, \au{Chowdhury, E.~A.}, \au{Galvanauskas, A.},
  \au{Gizzi, L.~A.}, \au{Hein, J.} \& \au{Hillier, D. I. et~al.}} \yr{2019}
  \at{{Petawatt and exawatt class lasers worldwide}}.  \jt{High Power Laser
  Sci.}  \bvol{7},  \pg{e54}.

\bibitem[Dorn {\em et~al.\/}(2003)Dorn, Quabis \& Leuchs]{2003_Dorn}
{\sc \au{Dorn, R.}, \au{Quabis, S.} \& \au{Leuchs, G.}} \yr{2003}  \at{{Sharper
  Focus for a Radially Polarized Light Beam}}.  \jt{Phys. Rev. Lett.}
  \bvol{91},  \pg{233901}.

\bibitem[Ehlotzky {\em et~al.\/}(2009)Ehlotzky, Krajewska \&
  Kami{\'{n}}ski]{2009_EhlotzkyF}
{\sc \au{Ehlotzky, F.}, \au{Krajewska, K.} \& \au{Kami{\'{n}}ski, J.~Z.}}
  \yr{2009}  \at{{Fundamental processes of quantum electrodynamics in laser
  fields of relativistic power}}.  \jt{Rep. Prog. Phys.}  \bvol{72}~(4),
  \pg{046401}.

\bibitem[Eliasson \& Liu(2013)]{2013_EliassonB}
{\sc \au{Eliasson, B.} \& \au{Liu, C.~S.}} \yr{2013}  \at{{An electromagnetic
  gamma-ray free electron laser}}.  \jt{J. Plasma Phys.}  \bvol{79}~(6),
  \pg{995–998}.

\bibitem[Esirkepov {\em et~al.\/}(2004)Esirkepov, Borghesi, Bulanov, Mourou \&
  Tajima]{2004_EsirkepovTZ}
{\sc \au{Esirkepov, T.~Z.}, \au{Borghesi, M.}, \au{Bulanov, S.~V.}, \au{Mourou,
  G.} \& \au{Tajima, T.}} \yr{2004}  \at{{Highly Efficient Relativistic-Ion
  Generation in the Laser-Piston Regime}}.  \jt{Phys. Rev. Lett.}  \bvol{92},
  \pg{175003}.

\bibitem[Ghotra \& Kant(2015)]{2015_Ghotra}
{\sc \au{Ghotra, H.~S.} \& \au{Kant, N.}} \yr{2015}  \at{{Sensitiveness of
  axial magnetic field on electron acceleration by a radially polarized laser
  pulse in vacuum}}.  \jt{Opt. Commun.}  \bvol{356},  \pg{118--122}.

\bibitem[Gong {\em et~al.\/}(2017)Gong, Hu, Shou, Qiao, Chen, He, Bulanov,
  Esirkepov, Bulanov \& Yan]{2017_GongZ}
{\sc \au{Gong, Z.}, \au{Hu, R.~H.}, \au{Shou, Y.~R.}, \au{Qiao, B.}, \au{Chen,
  C.~E.}, \au{He, X.~T.}, \au{Bulanov, S.~S.}, \au{Esirkepov, T.~Zh.},
  \au{Bulanov, S.~V.} \& \au{Yan, X.~Q.}} \yr{2017}  \at{{High-efficiency
  $\ensuremath{\gamma}$-ray flash generation via multiple-laser scattering in
  ponderomotive potential well}}.  \jt{Phys. Rev. E}  \bvol{95},  \pg{013210}.

\bibitem[Grismayer {\em et~al.\/}(2016)Grismayer, Vranic, Martins, Fonseca \&
  Silva]{2016_GrismayerT}
{\sc \au{Grismayer, T.}, \au{Vranic, M.}, \au{Martins, J.~L.}, \au{Fonseca,
  R.~A.} \& \au{Silva, L.~O.}} \yr{2016}  \at{{Laser absorption via quantum
  electrodynamics cascades in counter propagating laser pulses}}.  \jt{Phys.
  Plasmas}  \bvol{23}~(5),  \pg{056706}.

\bibitem[Gu {\em et~al.\/}(2018)Gu, Klimo, Bulanov \& Weber]{2018_GuYJ}
{\sc \au{Gu, Y.~J.}, \au{Klimo, O.}, \au{Bulanov, S.~V.} \& \au{Weber, S.}}
  \yr{2018}  \at{{Brilliant gamma-ray beam and electron–positron pair
  production by enhanced attosecond pulses}}.  \jt{Commun. Phys.}  \bvol{1},
  \pg{1--9}.

\bibitem[Hadjisolomou {\em et~al.\/}(2021)Hadjisolomou, Jeong, Valenta, Korn \&
  Bulanov]{2021_HadjisolomouP}
{\sc \au{Hadjisolomou, P.}, \au{Jeong, T.~M.}, \au{Valenta, P.}, \au{Korn, G.}
  \& \au{Bulanov, S.~V.}} \yr{2021}  \at{{Gamma-ray flash generation in
  irradiating a thin foil target by a single-cycle tightly focused extreme
  power laser pulse}}.  \jt{Phys. Rev. E}  \bvol{104},  \pg{015203}.

\bibitem[Hayward(1970)]{1970_HaywardE}
{\sc \au{Hayward, E}} \yr{1970} {\em {Photonuclear Reactions}\/}.
  \publ{Washington, D.C., USA: National Bureau of Standards}.

\bibitem[Higuera \& Cary(2017)]{2017_HigueraAV}
{\sc \au{Higuera, A.~V.} \& \au{Cary, J.~R.}} \yr{2017}
  \at{{Structure-preserving second-order integration of relativistic charged
  particle trajectories in electromagnetic fields}}.  \jt{Phys. Plasmas}
  \bvol{24}~(5),  \pg{052104}.

\bibitem[Ilderton(2019)]{2019_IldertonA}
{\sc \au{Ilderton, A.}} \yr{2019}  \at{{Note on the conjectured breakdown of
  QED perturbation theory in strong fields}}.  \jt{Phys. Rev. D}  \bvol{99},
  \pg{085002}.

\bibitem[Jeong {\em et~al.\/}(2018)Jeong, Bulanov, Weber \& Korn]{2018_JeongTM}
{\sc \au{Jeong, T.~M.}, \au{Bulanov, S.~V.}, \au{Weber, S.} \& \au{Korn, G.}}
  \yr{2018}  \at{{Analysis on the longitudinal field strength formed by
  tightly-focused radially-polarized femtosecond petawatt laser pulse}}.
  \jt{Opt. Express}  \bvol{26}~(25),  \pg{33091--33107}.

\bibitem[Jeong {\em et~al.\/}(2015)Jeong, Weber, Le~Garrec, Margarone, Mocek \&
  Korn]{2015_JeongTM}
{\sc \au{Jeong, T.~M.}, \au{Weber, S.}, \au{Le~Garrec, B.}, \au{Margarone, D.},
  \au{Mocek, T.} \& \au{Korn, G.}} \yr{2015}  \at{{Spatio-temporal modification
  of femtosecond focal spot under tight focusing condition}}.  \jt{Opt.
  Express}  \bvol{23}~(9),  \pg{11641--11656}.

\bibitem[Ji {\em et~al.\/}(2019)Ji, Snyder \& Shen]{2019_JiLL}
{\sc \au{Ji, L.~L.}, \au{Snyder, J.} \& \au{Shen, B.~F.}} \yr{2019}
  \at{{Single-pulse laser-electron collision within a micro-channel plasma
  target}}.  \jt{Plasma Phys. Control. Fusion}  \bvol{61}~(6),  \pg{065019}.

\bibitem[Kirk {\em et~al.\/}(2009)Kirk, Bell \& Arka]{2009_KirkJG}
{\sc \au{Kirk, J.~G.}, \au{Bell, A.~R.} \& \au{Arka, I.}} \yr{2009}  \at{{Pair
  production in counter-propagating laser beams}}.  \jt{Plasma Phys. Control.
  Fusion}  \bvol{51}~(8),  \pg{085008}.

\bibitem[Klimo {\em et~al.\/}(2008)Klimo, Psikal, Limpouch \&
  Tikhonchuk]{2008_KlimoO}
{\sc \au{Klimo, O.}, \au{Psikal, J.}, \au{Limpouch, J.} \& \au{Tikhonchuk,
  V.~T.}} \yr{2008}  \at{{Monoenergetic ion beams from ultrathin foils
  irradiated by ultrahigh-contrast circularly polarized laser pulses}}.
  \jt{Phys. Rev. ST Accel. Beams}  \bvol{11},  \pg{031301}.

\bibitem[Koch \& Motz(1959)]{1959_KochHW}
{\sc \au{Koch, H.~W.} \& \au{Motz, J.~W.}} \yr{1959}  \at{{Bremsstrahlung
  Cross-Section Formulas and Related Data}}.  \jt{Rev. Mod. Phys.}  \bvol{31},
  \pg{920--955}.

\bibitem[Koga {\em et~al.\/}(2005)Koga, Esirkepov \& Bulanov]{2005_KogaJ}
{\sc \au{Koga, J.}, \au{Esirkepov, T.~Z.} \& \au{Bulanov, S.~V.}} \yr{2005}
  \at{{Nonlinear Thomson scattering in the strong radiation damping regime}}.
  \jt{Phys Plasmas}  \bvol{12}~(9),  \pg{093106}.

\bibitem[Landau(1944)]{1944_LandauL}
{\sc \au{Landau, L.}} \yr{1944}  \at{{On the energy loss of fast particles by
  ionization}}.  \jt{J. Phys. USSR}  \bvol{8}~(1-6),  \pg{201--205}.

\bibitem[Lezhnin {\em et~al.\/}(2018)Lezhnin, Sasorov, Korn \&
  Bulanov]{2018_LezhninKV}
{\sc \au{Lezhnin, K.~V.}, \au{Sasorov, P.~V.}, \au{Korn, G.} \& \au{Bulanov,
  S.~V.}} \yr{2018}  \at{{High power gamma flare generation in multi-petawatt
  laser interaction with tailored targets}}.  \jt{Phys. Plasmas}
  \bvol{25}~(12),  \pg{123105}.

\bibitem[Li {\em et~al.\/}(2012)Li, Salamin, Galow \& Keitel]{2012_JianXing}
{\sc \au{Li, J.~X.}, \au{Salamin, Y.~I.}, \au{Galow, B.~J.} \& \au{Keitel,
  C.~H.}} \yr{2012}  \at{{Acceleration of proton bunches by petawatt chirped
  radially polarized laser pulses}}.  \jt{Phys. Rev. A}  \bvol{85},
  \pg{063832}.

\bibitem[Li {\em et~al.\/}(2021)Li, Kato \& Kawanaka]{2021_LiZ}
{\sc \au{Li, Z.}, \au{Kato, Y.} \& \au{Kawanaka, J.}} \yr{2021}
  \at{{Simulating an ultra-broadband concept for Exawatt-class lasers}}.
  \jt{Sci. Rep.}  \bvol{11}~(151).

\bibitem[Luo {\em et~al.\/}(2015)Luo, Zhu, Zhuo, Ma, Song, Zhu, Wang, Li, Turcu
  \& Chen]{2015_LuoW}
{\sc \au{Luo, W.}, \au{Zhu, Y.~B.}, \au{Zhuo, H.~B.}, \au{Ma, Y.~Y.}, \au{Song,
  Y.~M.}, \au{Zhu, Z.~C.}, \au{Wang, X.~D.}, \au{Li, X.~H.}, \au{Turcu, I.
  C.~E.} \& \au{Chen, M.}} \yr{2015}  \at{{Dense electron-positron plasmas and
  gamma-ray bursts generation by counter-propagating quantum
  electrodynamics-strong laser interaction with solid targets}}.  \jt{Physics
  of Plasmas}  \bvol{22}~(6),  \pg{063112}.

\bibitem[Magnusson {\em et~al.\/}(2019)Magnusson, Gonoskov, Marklund,
  Esirkepov, Koga, Kondo, Kando, Bulanov, Korn \& Bulanov]{2019_MagnussonJ}
{\sc \au{Magnusson, J.}, \au{Gonoskov, A.}, \au{Marklund, M.}, \au{Esirkepov,
  T.~Z.}, \au{Koga, J.~K.}, \au{Kondo, K.}, \au{Kando, M.}, \au{Bulanov,
  S.~V.}, \au{Korn, G.} \& \au{Bulanov, S.~S.}} \yr{2019}  \at{{Laser-Particle
  Collider for Multi-GeV Photon Production}}.  \jt{Phys. Rev. Lett.}
  \bvol{122},  \pg{254801}.

\bibitem[Mourou {\em et~al.\/}(2002)Mourou, Chang, Maksimchuk, Nees, Bulanov,
  Bychenkov, Esirkepov, Naumova, Pegoraro \& Ruhl]{2002_MourouG}
{\sc \au{Mourou, G.}, \au{Chang, Z.}, \au{Maksimchuk, A.}, \au{Nees, J.},
  \au{Bulanov, S.~V.}, \au{Bychenkov, V.~Y.}, \au{Esirkepov, T.~Z.},
  \au{Naumova, N.~M.}, \au{Pegoraro, F.} \& \au{Ruhl, H}} \yr{2002}  \at{{On
  the design of experiments for the study of relativistic nonlinear optics in
  the limit of single-cycle pulse duration and single-wavelength spot size}}.
  \jt{Plasma Phys. Rep.}  \bvol{28},  \pg{12--27}.

\bibitem[Mourou {\em et~al.\/}(2006)Mourou, Tajima \& Bulanov]{2006_MourouG}
{\sc \au{Mourou, G.~A.}, \au{Tajima, T.} \& \au{Bulanov, S.~V.}} \yr{2006}
  \at{{Optics in the relativistic regime}}.  \jt{Rev. Mod. Phys.}  \bvol{78},
  \pg{309--371}.

\bibitem[Nakamura {\em et~al.\/}(2012)Nakamura, Koga, Esirkepov, Kando, Korn \&
  Bulanov]{2012_NakamuraT}
{\sc \au{Nakamura, T.}, \au{Koga, J.~K.}, \au{Esirkepov, T.~Z.}, \au{Kando,
  M.}, \au{Korn, G.} \& \au{Bulanov, S.~V.}} \yr{2012}  \at{{High-Power
  $\ensuremath{\gamma}$-Ray Flash Generation in Ultraintense Laser-Plasma
  Interactions}}.  \jt{Phys. Rev. Lett.}  \bvol{108},  \pg{195001}.

\bibitem[Narozhny(1979)]{1979_NarozhnyNB}
{\sc \au{Narozhny, N.~B.}} \yr{1979}  \at{{Radiation corrections to quantum
  processes in an intense electromagnetic field}}.  \jt{Phys. Rev. D}
  \bvol{20},  \pg{1313--1320}.

\bibitem[Nedorezov {\em et~al.\/}(2004)Nedorezov, Turinge \&
  Shatunov]{2004_NedorezovVG}
{\sc \au{Nedorezov, V.~G.}, \au{Turinge, A.~A.} \& \au{Shatunov, Y.~M.}}
  \yr{2004}  \at{{Photonuclear experiments with Compton-backscattered gamma
  beams}}.  \jt{Phys.-Uspekhi}  \bvol{47}~(4),  \pg{341--358}.

\bibitem[Osvay {\em et~al.\/}(2019)Osvay, Börzsönyi, Cao, Cormier, Csontos,
  Jójárt, Kalashnikov, Kiss, López-Martens \& Tóth]{2019_Osvay}
{\sc \au{Osvay, K.}, \au{Börzsönyi, A.}, \au{Cao, H.}, \au{Cormier, E.},
  \au{Csontos, J.}, \au{Jójárt, P.}, \au{Kalashnikov, M.}, \au{Kiss, B.},
  \au{López-Martens, r.} \& \au{Tóth, S. et~al.}} \yr{2019} {Development
  status and operation experiences of the few cycle high average power lasers
  of ELI-ALPS (Conference Presentation)}.  \bt{In {\em Short-pulse High-energy
  Lasers and Ultrafast Optical Technologies\/} (ed. \ed{Pavel Bakule \&
  Constantin~L. Haefner})}, ,  \vol{vol. 11034}. International Society for
  Optics and Photonics,  \publ{SPIE}.

\bibitem[Ouillé {\em et~al.\/}(2020)Ouillé, Vernier, Böhle, Bocoum, Jullien,
  Lozano, Rousseau, Cheng, Gustas \& Blumenstein]{2020_OuilleM}
{\sc \au{Ouillé, M.}, \au{Vernier, A.}, \au{Böhle, F.}, \au{Bocoum, M.},
  \au{Jullien, A.}, \au{Lozano, M.}, \au{Rousseau, J.~P.}, \au{Cheng, Z.},
  \au{Gustas, D.} \& \au{Blumenstein, A. et~al.}} \yr{2020}
  \at{{Relativistic-intensity near-single-cycle light waveforms at kHz
  repetition rate}}.  \jt{Light Sci. Appl.}  \bvol{9}.

\bibitem[Payeur {\em et~al.\/}(2012)Payeur, Fourmaux, Schmidt, MacLean,
  Tchervenkov, Légaré, Piché \& Kieffer]{2012_Payeur}
{\sc \au{Payeur, S.}, \au{Fourmaux, S.}, \au{Schmidt, B.~E.}, \au{MacLean,
  J.~P.}, \au{Tchervenkov, C.}, \au{Légaré, F.}, \au{Piché, M.} \&
  \au{Kieffer, J.~C.}} \yr{2012}  \at{{Generation of a beam of fast electrons
  by tightly focusing a radially polarized ultrashort laser pulse}}.  \jt{Appl.
  Phys. Lett.}  \bvol{101}~(4),  \pg{041105}.

\bibitem[Perry {\em et~al.\/}(1999)Perry, Pennington, Stuart, Tietbohl,
  Britten, Brown, Herman, Golick, Kartz, Miller \& et~al.]{1999_Perry}
{\sc \au{Perry, M.~D.}, \au{Pennington, D.}, \au{Stuart, B.~C.}, \au{Tietbohl,
  G.}, \au{Britten, J.~A.}, \au{Brown, C.}, \au{Herman, S.}, \au{Golick, B.},
  \au{Kartz, M.}, \au{Miller, J.} \& \au{et~al.}} \yr{1999}  \at{{Petawatt
  laser pulses}}.  \jt{Opt. Lett.}  \bvol{24}~(3),  \pg{160--162}.

\bibitem[Philippov \& Spitkovsky(2018)]{2018_PhilippovAA}
{\sc \au{Philippov, A.~A.} \& \au{Spitkovsky, A.}} \yr{2018}  \at{{Ab-initio
  Pulsar Magnetosphere: Particle Acceleration in Oblique Rotators and
  High-energy Emission Modeling}}.  \jt{Astrophys. J.}  \bvol{855}~(2),
  \pg{94}.

\bibitem[Pirozhkov {\em et~al.\/}(2017)Pirozhkov, Fukuda, Nishiuchi, Kiriyama,
  Sagisaka, Ogura, Mori, Kishimoto, Sakaki \& Dover]{2017_PirozhkovAS}
{\sc \au{Pirozhkov, A.~S.}, \au{Fukuda, Y.}, \au{Nishiuchi, M.}, \au{Kiriyama,
  H.}, \au{Sagisaka, A.}, \au{Ogura, K.}, \au{Mori, M.}, \au{Kishimoto, M.},
  \au{Sakaki, H.} \& \au{Dover, N. P. et~al.}} \yr{2017}  \at{{Approaching the
  diffraction-limited, bandwidth-limited Petawatt}}.  \jt{Opt. Express}
  \bvol{25}~(17),  \pg{20486--20501}.

\bibitem[Rees \& Mészáros(1992)]{1992_ReesMJ}
{\sc \au{Rees, M.~J.} \& \au{Mészáros, P.}} \yr{1992}  \at{{Relativistic
  fireballs: energy conversion and time-scales}}.  \jt{Mon. Not. R. Astron.
  Soc.}  \bvol{258}~(1),  \pg{41P--43P}.

\bibitem[Richards {\em et~al.\/}(1959)Richards, Wolf \& Gabor]{1959_RichardsB}
{\sc \au{Richards, B.}, \au{Wolf, E.} \& \au{Gabor, D.}} \yr{1959}
  \at{{Electromagnetic diffraction in optical systems, II. Structure of the
  image field in an aplanatic system}}.  \jt{Proceedings of the Royal Society
  of London. Series A. Mathematical and Physical Sciences}  \bvol{253}~(1274),
  \pg{358--379}.

\bibitem[Ridgers {\em et~al.\/}(2013)Ridgers, Brady, Duclous, Kirk, Bennett,
  Arber \& Bell]{2013_RidgersCP}
{\sc \au{Ridgers, C.~P.}, \au{Brady, C.~S.}, \au{Duclous, R.}, \au{Kirk,
  J.~G.}, \au{Bennett, K.}, \au{Arber, T.~D.} \& \au{Bell, A.~R.}} \yr{2013}
  \at{{Dense electron-positron plasmas and bursts of gamma-rays from
  laser-generated quantum electrodynamic plasmas}}.  \jt{Phys. Plasmas}
  \bvol{20}~(5),  \pg{056701}.

\bibitem[Ridgers {\em et~al.\/}(2012)Ridgers, Brady, Duclous, Kirk, Bennett,
  Arber, Robinson \& Bell]{2012_RidgersCP}
{\sc \au{Ridgers, C.~P.}, \au{Brady, C.~S.}, \au{Duclous, R.}, \au{Kirk,
  J.~G.}, \au{Bennett, K.}, \au{Arber, T.~D.}, \au{Robinson, A. P.~L.} \&
  \au{Bell, A.~R.}} \yr{2012}  \at{{Dense Electron-Positron Plasmas and
  Ultraintense $\ensuremath{\gamma}$ rays from Laser-Irradiated Solids}}.
  \jt{Phys. Rev. Lett.}  \bvol{108},  \pg{165006}.

\bibitem[Ridgers {\em et~al.\/}(2014)Ridgers, Kirk, Duclous, Blackburn, Brady,
  Bennett, Arber \& Bell]{2014_RidgersCP}
{\sc \au{Ridgers, C.~P.}, \au{Kirk, J.~G.}, \au{Duclous, R.}, \au{Blackburn,
  T.~G.}, \au{Brady, C.~S.}, \au{Bennett, K.}, \au{Arber, T.~D.} \& \au{Bell,
  A.~R.}} \yr{2014}  \at{{Modelling gamma-ray photon emission and pair
  production in high-intensity laser–matter interactions}}.  \jt{J. Comput.
  Phys}  \bvol{260},  \pg{273--285}.

\bibitem[Ritus(1970)]{1970_RitusVI}
{\sc \au{Ritus, V.~I.}} \yr{1970}  \at{{Radiative Effects and Their Enhancement
  in an Intense Electromagnetic Field}}.  \jt{J. Exp. Theor. Phys.}
  \bvol{30}~(6),  \pg{1181}.

\bibitem[Rivas {\em et~al.\/}(2017)Rivas, Borot, Cardenas, Marcus, Gu,
  Herrmann, Xu, Tan, Kormin \& Ma]{2017_Rivas}
{\sc \au{Rivas, D.~E.}, \au{Borot, A.}, \au{Cardenas, D.~E.}, \au{Marcus, G.},
  \au{Gu, X.}, \au{Herrmann, D.}, \au{Xu, J.}, \au{Tan, J.}, \au{Kormin, D.} \&
  \au{Ma, G. et~al.}} \yr{2017}  \at{{Next Generation Driver for Attosecond and
  Laser-plasma Physics}}.  \jt{Sci. Rep.}  \bvol{7}.

\bibitem[Robinson {\em et~al.\/}(2008)Robinson, Zepf, Kar, Evans \&
  Bellei]{2008_RobinsonAPL}
{\sc \au{Robinson, A. P.~L.}, \au{Zepf, M.}, \au{Kar, S.}, \au{Evans, R.~G.} \&
  \au{Bellei, C.}} \yr{2008}  \at{{Radiation pressure acceleration of thin
  foils with circularly polarized laser pulses}}.  \jt{New J. Phys.}
  \bvol{10}~(1),  \pg{013021}.

\bibitem[Salamin(2006)]{2006_Salamin}
{\sc \au{Salamin, Y.~I.}} \yr{2006}  \at{{Fields of a radially polarized
  Gaussian laser beam beyond the paraxial approximation}}.  \jt{Opt. Lett.}
  \bvol{31}~(17),  \pg{2619--2621}.

\bibitem[Salamin(2010{\natexlab{{\em a\/}}})]{2010_Salamin}
{\sc \au{Salamin, Y.~I.}} \yr{2010{\natexlab{{\em a\/}}}}  \at{{Direct particle
  acceleration by two identical crossed radially polarized laser beams}}.
  \jt{Phys. Rev. A}  \bvol{82},  \pg{013823}.

\bibitem[Salamin(2010{\natexlab{{\em b\/}}})]{2010b_Salamin}
{\sc \au{Salamin, Y.~I.}} \yr{2010{\natexlab{{\em b\/}}}}  \at{{Low-diffraction
  direct particle acceleration by a radially polarized laser beam}}.  \jt{Phys.
  Lett. A}  \bvol{374}~(48),  \pg{4950--4953}.

\bibitem[Salamin(2015)]{2015_Salamin}
{\sc \au{Salamin, Y.~I.}} \yr{2015}  \at{{Fields and propagation
  characteristics in vacuum of an ultrashort tightly focused radially polarized
  laser pulse}}.  \jt{Phys. Rev. A}  \bvol{92},  \pg{053836}.

\bibitem[Sales(1998)]{1998_Sales}
{\sc \au{Sales, T. R.~M.}} \yr{1998}  \at{{Smallest Focal Spot}}.  \jt{Phys.
  Rev. Lett.}  \bvol{81},  \pg{3844--3847}.

\bibitem[Sarri {\em et~al.\/}(2015)Sarri, Poder, Cole, Schumaker, Di~Piazza,
  Reville, Dzelzainis, Doria, Gizzi \& Grittani]{2015_SarriG}
{\sc \au{Sarri, G.}, \au{Poder, K.}, \au{Cole, J.~M.}, \au{Schumaker, W.},
  \au{Di~Piazza, A.}, \au{Reville, B.}, \au{Dzelzainis, T.}, \au{Doria, D.},
  \au{Gizzi, L.~A.} \& \au{Grittani, G. et~al.}} \yr{2015}  \at{{Generation of
  neutral and high-density electron--positron pair plasmas in the laboratory}}.
   \jt{Nat. Commun.}  \bvol{6}~(6747),  \pg{1--8}.

\bibitem[Schneider {\em et~al.\/}(2002)Schneider, Agosteo, Pedroni \&
  Besserer]{2002_SchneiderU}
{\sc \au{Schneider, U.}, \au{Agosteo, S.}, \au{Pedroni, E.} \& \au{Besserer,
  J.}} \yr{2002}  \at{{Secondary neutron dose during proton therapy using spot
  scanning}}.  \jt{Int. J. Radiat. Oncol. Biol. Phys.}  \bvol{53}~(1),
  \pg{244--251}.

\bibitem[Strickland \& Mourou(1985)]{1985_StricklandD}
{\sc \au{Strickland, D.} \& \au{Mourou, G.}} \yr{1985}  \at{{Compression of
  amplified chirped optical pulses}}.  \jt{Opt. Commun.}  \bvol{56}~(3),
  \pg{219--221}.

\bibitem[Tadamura {\em et~al.\/}(1999)Tadamura, Kudoh, Motooka, Inubushi,
  Shirakawa, Hattori, Okada, Matsuda, Koshiji \& Nishimura]{1999_TadamuraE}
{\sc \au{Tadamura, E.}, \au{Kudoh, T.}, \au{Motooka, M.}, \au{Inubushi, M.},
  \au{Shirakawa, S.}, \au{Hattori, N.}, \au{Okada, T.}, \au{Matsuda, T.},
  \au{Koshiji, T.} \& \au{Nishimura, K. et~al.}} \yr{1999}  \at{{Assessment of
  regional and global left ventricular function by reinjection Tl-201 and rest
  Tc-99m sestamibi ECG-gated SPECT: Comparison with three-dimensional magnetic
  resonance imaging}}.  \jt{J. Am. Coll. Cardiol.}  \bvol{33}~(4),
  \pg{991--997}.

\bibitem[Tanaka {\em et~al.\/}(2020)Tanaka, Spohr, Balabanski, Balascuta,
  Capponi, Cernaianu, Cuciuc, Cucoanes, Dancus \& Dhal]{2020_TanakaKA}
{\sc \au{Tanaka, K.~A.}, \au{Spohr, K.~M.}, \au{Balabanski, D.~L.},
  \au{Balascuta, S.}, \au{Capponi, L.}, \au{Cernaianu, M.~O.}, \au{Cuciuc, M.},
  \au{Cucoanes, A.}, \au{Dancus, I.} \& \au{Dhal, A. et~al.}} \yr{2020}
  \at{{Current status and highlights of the ELI-NP research program}}.
  \jt{Matter Radiat. at Extremes}  \bvol{5}~(2),  \pg{024402}.

\bibitem[Vlachoudis(2009)]{2009_VlachoudisV}
{\sc \au{Vlachoudis, V.}} \yr{2009} {FLAIR: a powerful but user friendly
  graphical interface for FLUKA}.  \bt{In {\em Proc. Int. Conf. on Mathematics,
  Computational Methods \& Reactor Physics (M\&C 2009), Saratoga Springs, New
  York\/}}, ,  \vol{vol. 176}.

\bibitem[{Voronin, A. A. and Zheltikov, A. M. and Ditmire, T. and Rus, B. and
  Korn, G.}(2013)]{2013_VoroninAA}
{\sc \au{{Voronin, A. A. and Zheltikov, A. M. and Ditmire, T. and Rus, B. and
  Korn, G.}}} \yr{2013}  \at{Subexawatt few-cycle lightwave generation via
  multipetawatt pulse compression}.  \jt{Opt. Commun.}  \bvol{291},
  \pg{299--303}.

\bibitem[Vranic {\em et~al.\/}(2016)Vranic, Grismayer, Fonseca \&
  Silva]{2016_VranicM}
{\sc \au{Vranic, M.}, \au{Grismayer, T.}, \au{Fonseca, R.~A.} \& \au{Silva,
  L.~O.}} \yr{2016}  \at{{Electron{\textendash}positron cascades in
  multiple-laser optical traps}}.  \jt{Plasma Phys. Control. Fusion}
  \bvol{59}~(1),  \pg{014040}.

\bibitem[Vshivkov {\em et~al.\/}(1998)Vshivkov, Naumova, Pegoraro \&
  Bulanov]{1998_VshivkovVA}
{\sc \au{Vshivkov, V.~A.}, \au{Naumova, N.~M.}, \au{Pegoraro, F.} \&
  \au{Bulanov, S.~V.}} \yr{1998}  \at{{Nonlinear electrodynamics of the
  interaction of ultra-intense laser pulses with a thin foil}}.  \jt{Phys.
  Plasmas}  \bvol{5}~(7),  \pg{2727--2741}.

\bibitem[Wang {\em et~al.\/}(2001)Wang, Ho, Yuan, Kong, Cao, Sessler, Esarey \&
  Nishida]{Wang2001}
{\sc \au{Wang, P.~X.}, \au{Ho, Y.~K.}, \au{Yuan, X.~Q.}, \au{Kong, Q.},
  \au{Cao, N.}, \au{Sessler, A.~M.}, \au{Esarey, E.} \& \au{Nishida, Y.}}
  \yr{2001}  \at{{Vacuum electron acceleration by an intense laser}}.
  \jt{Appl. Phys. Lett.}  \bvol{78}~(15),  \pg{2253--2255}.

\bibitem[Wang {\em et~al.\/}(2020)Wang, Hu, Zhang, Gu, Zhao, Zuo \&
  Zheng]{2020_WangXB}
{\sc \au{Wang, X.~B.}, \au{Hu, G.~Y.}, \au{Zhang, Z.~M.}, \au{Gu, Y.~Q.},
  \au{Zhao, B.}, \au{Zuo, Y.} \& \au{Zheng, J.}} \yr{2020}  \at{{Gamma-ray
  generation from ultraintense laser-irradiated solid targets with preplasma}}.
   \jt{High Power Laser Sci. Eng.}  \bvol{8},  \pg{e34}.

\bibitem[West(1982)]{1982_WestD}
{\sc \au{West, D.}} \at{ \yr{1982} } \bt{In {\em {Ternary Equilibrium
  Diagrams}\/}}.  \publ{Springer}.

\bibitem[Wheeler \& Lamb~Jr(1939)]{1939_WheelerJA}
{\sc \au{Wheeler, J.~A.} \& \au{Lamb~Jr, W.E.}} \yr{1939}  \at{{Influence of
  atomic electrons on radiation and pair production}}.  \jt{Phys. Rev.}
  \bvol{55}~(9),  \pg{858}.

\bibitem[Younis {\em et~al.\/}(2021)Younis, Davidson, Hafizi \&
  Gordon]{2021_YounisAH}
{\sc \au{Younis, A.~H.}, \au{Davidson, A.}, \au{Hafizi, B.} \& \au{Gordon,
  D.~F.}} \yr{2021} {Diagnostic Techniques for Particle-in-Cell Simulations of
  Laser-produced Gamma-rays in the Strong-field QED Regime},  \arxiv{arXiv:
  2106.16227}.

\bibitem[Zhang {\em et~al.\/}(2021)Zhang, Wu, Huang, Lan, Liu, Wu, Yang, Zhao,
  Zhu \& Luo]{2021_ZhangLQ}
{\sc \au{Zhang, L.~Q.}, \au{Wu, S.~D.}, \au{Huang, H.~R.}, \au{Lan, H.~Y.},
  \au{Liu, W.~Y.}, \au{Wu, Y.~C.}, \au{Yang, Y.}, \au{Zhao, Z.~Q.}, \au{Zhu,
  C.~H.} \& \au{Luo, W.}} \yr{2021}  \at{{Brilliant attosecond \textgamma-ray
  emission and high-yield positron production from intense laser-irradiated
  nano-micro array}}.  \jt{Phys. Plasmas}  \bvol{28}~(2),  \pg{023110}.

\bibitem[Zhidkov {\em et~al.\/}(2002)Zhidkov, Koga, Sasaki \&
  Uesaka]{2002_ZhidkovA}
{\sc \au{Zhidkov, A.}, \au{Koga, J.}, \au{Sasaki, A.} \& \au{Uesaka, M.}}
  \yr{2002}  \at{{Radiation Damping Effects on the Interaction of Ultraintense
  Laser Pulses with an Overdense Plasma}}.  \jt{Phys. Rev. Lett.}  \bvol{88},
  \pg{185002}.

\end{thebibliography}

\end{document}